\documentclass{ws-m3as}

\usepackage{amscd}
\begin{document}

\markboth{C. Arita, A. Schadschneider}{
EXCLUSIVE QUEUEING PROCESSES AND 
THEIR APPLICATION TO TRAFFIC SYSTEMS}

%%%%%%%%%%%%%%%%%%% Publisher's Area please ignore %%%%%%%%%%%%%%%%%%%%%%%
%
\catchline{}{}{}{}{}
%
%%%%%%%%%%%%%%%%%%%%%%%%%%%%%%%%%%%%%%%%%%%%%%%%%%%%%%%%%%%%%%%%%%%%%%%%%%

\title{EXCLUSIVE QUEUEING PROCESSES
AND THEIR APPLICATION TO TRAFFIC SYSTEMS}

\author{CHIKASHI ARITA}

\address{Theoretische Physik, Universit\"{a}t des Saarlandes\\
66041 Saarbr\"ucken, Germany\\
%\footnote{State completely without abbreviations, the affiliation and 
%mailing address, including country. Typeset in 8 pt Times italic.}\\
c.arita@physik.uni-saarland.de}

\author{ANDREAS SCHADSCHNEIDER}

\address{Institut f\"ur Theoretische Physik, Universit\"at zu K\"oln\\
50937 K\"oln, Germany\\
as@thp.uni-koeln.de}

\maketitle

\begin{history}
\received{\today}
%(Day Month Year)}
\revised{(Day Month Year)}
%\accepted{(Day Month Year)}
\comby{(xxxxxxxxxx)}
\end{history}

\begin{abstract}
  The dynamics of pedestrian crowds has been studied intensively in
  recent years, both theoretically and empirically.  However, in many
  situations pedestrian crowds are rather static, e.g. due to jamming
  near bottlenecks or queueing at ticket counters or supermarket
  checkouts.  Classically such queues are often described by the M/M/1
  queue that neglects the internal structure (density profile) of the
  queue by focussing on the system length as the only dynamical
  variable. This is different in the Exclusive Queueing Process (EQP)
  in which the queue is considered on a microscopic level. It is
  equivalent to a Totally Asymmetric Exclusion Process (TASEP) of
  varying length.  The EQP has a surprisingly rich phase diagram with
  respect to the arrival probability $\alpha$ and the service
  probability $\beta$.  The behavior on the phase transition line is
  much more complex than for the TASEP with a fixed system length.  It
  is nonuniversal and depends strongly on the update procedure used.
  In this article, we review the main properties of the EQP. 
  We also mention extensions and applications of the EQP
  and some related models.  
\end{abstract}


%22E46, 53C35, 57S20
%60K25 (queueing theory)
%82C22 (Statistical mechanics: interaction particle systems)
%90B22 (queues and service)
%90B20 (traffic)
%37F99 (Dynamical system and ergodic theory; none of the above)
%37E20 (Low dimensional systems; universality, renormalization)
%%%%%%%%%%%%%%%%%%%%%%%%%%%%%%%%%%%%%%%%%%%%%%%%%%%%%%%%%%%%%%%%%%%%%%%%%%%%%

\section{Introduction}    

\label{sec:intro}

Queueing processes have been studied extensively for decades
\cite{bib:Medhi,bib:Saaty}. Although originally developed to describe
problems of telecommunication, they have been applied later also to
various kinds of jamming phenomena, e.g.\ supply chains 
and vehicular traffic (see Sec.~\ref{sec-applications}).
However, classical queueing theory neglects the spatial structure of 
queues and the customers (particles) in queues do not interact with each other.
The length $L_t$ of the system  is the only dynamical variable and the
density along the queue is constant.
Therefore an extension of the classical M/M/1 queueing process
has been introduced recently \cite{bib:A,bib:Y}. It takes into account 
particle interactions through the excluded-volume effect and leads
to nontrivial density profiles of the queue.

Classical queueing theory has been introduced more than 100 years ago
with the seminal works by Erlang \cite{erlang}. It is closely related
to the theory of Markov chains and has found many applications ranging
from telecommunication and traffic flow to economics.  Queueing models
are usually classified according to the type of the arrival processes,
the service time distribution and the number of queues, which are
denoted by Kendall's notation.  The queue discipline (e.g.
First-In-First-Out (FIFO) or Last-In-First-Out (LIFO)) is also an
important classification.

%%%%%%%%%%%%%%%%%%%%%%%%%%%%%%%%%%%%%%%%%%%%%%%%%%%%%%%%%%%%%%%%%%%%%%%%%%%%%

\section{Markov chains and classical queueing theory}

Markov chains (or Markov processes) \cite{bib:Schi} have become an
important tool to phenomenologically describe physical systems
\cite{bib:L,bib:Schue,bib:ZS,bib:SCN,bib:KRB}.  The dynamics of a
Markov chain with discrete time $t$ on a state space $S$, which is a
countable set, is governed by
\begin{equation}
P(\tau;t+1) = \sum_{\tau'\in S} W(\tau'\to\tau) P(\tau';t)  , 
\label{eq:master}
\end{equation}   
where $W(\tau'\to\tau)$ is the transition probability from $\tau'$ to
$\tau$ \footnote{Here we assume that the transition probability is
  independent of $t$. }, and $ P(\tau ;t) \ (\tau\in S )$ is the
probability of finding the state $\tau$ at time $t$. Physicists often
call this equation ``master equation'' \cite{bib:Schue}.  When we
achieve any $\tau_f\in S $ from any $\tau_i\in S $ (i.e. there is a
path $\tau_i\to\tau_1\to \cdots \tau_n\to \tau_f $ such that
$W(\tau_i\to \tau_1)W(\tau_1\to \tau_2) \dots W(\tau_n\to \tau_f)>0
$), we say that the system is irreducible.

The (a)periodicity is also of importance for the Markov processes.
The period of a state $ \tau \in S $ is defined as
gcd$ \{ n | W(\tau\to\tau_1)\cdots W(\tau_n\to\tau) >0 \}$ 
(greatest common divisor). 
For an irreducible Markov process, all the states have the same
period. When the period is 1, we say the process is aperiodic. Note
that if a process has at least one state $\tau$ such that $ W(\tau\to
\tau) >0 $, the process is aperiodic.

The stationary measure is the solution to\footnote{
Unnormalizable stationary measures are not always unique. }
\begin{equation}
P_{st}(\tau ) = \sum_{\tau'\in S} W(\tau'\to\tau) F(\tau' )  .
\label{eq:stationary}
\end{equation}
When a stationary measure $P_{st}(\tau ) $ is normalizable, i.e.
$\sum_{\tau\in S}P_{st} (\tau):=Z$ is finite, we can construct a
stationary distribution by $\frac 1 Z P_{st}(\tau)$. For an
irreducible and aperiodic system, a stationary distribution is unique,
if it exists, and we have the important property $\lim_{t\to\infty}
P(\tau;t) = \frac 1 Z P_{st}(\tau)$ \cite{bib:Schi} \footnote{ This
  can be proved by the Perron-Frobenius theorem when $S$ is a finite
  set, although the proof becomes complicated when $S$ is infinite
  (but countable) set.}.

When a stationary distribution does not exist, 
we have $\lim_{t\to\infty} P(\tau;t) = 0$  for all $\tau\in S$.

%%%%%%%%%%%%%%%%%%%%%%%%%%%%%%%%%%%%%%%%%%%%%%%%%%%%

\subsection{M/M/1 queue}

\begin{figure}[t]
\begin{center}
  \includegraphics[width=0.35\textwidth]{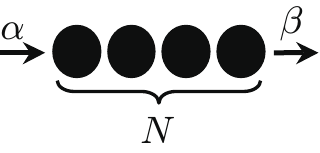}
  \qquad\quad\quad\quad\quad
  \includegraphics[width=0.2\textwidth]{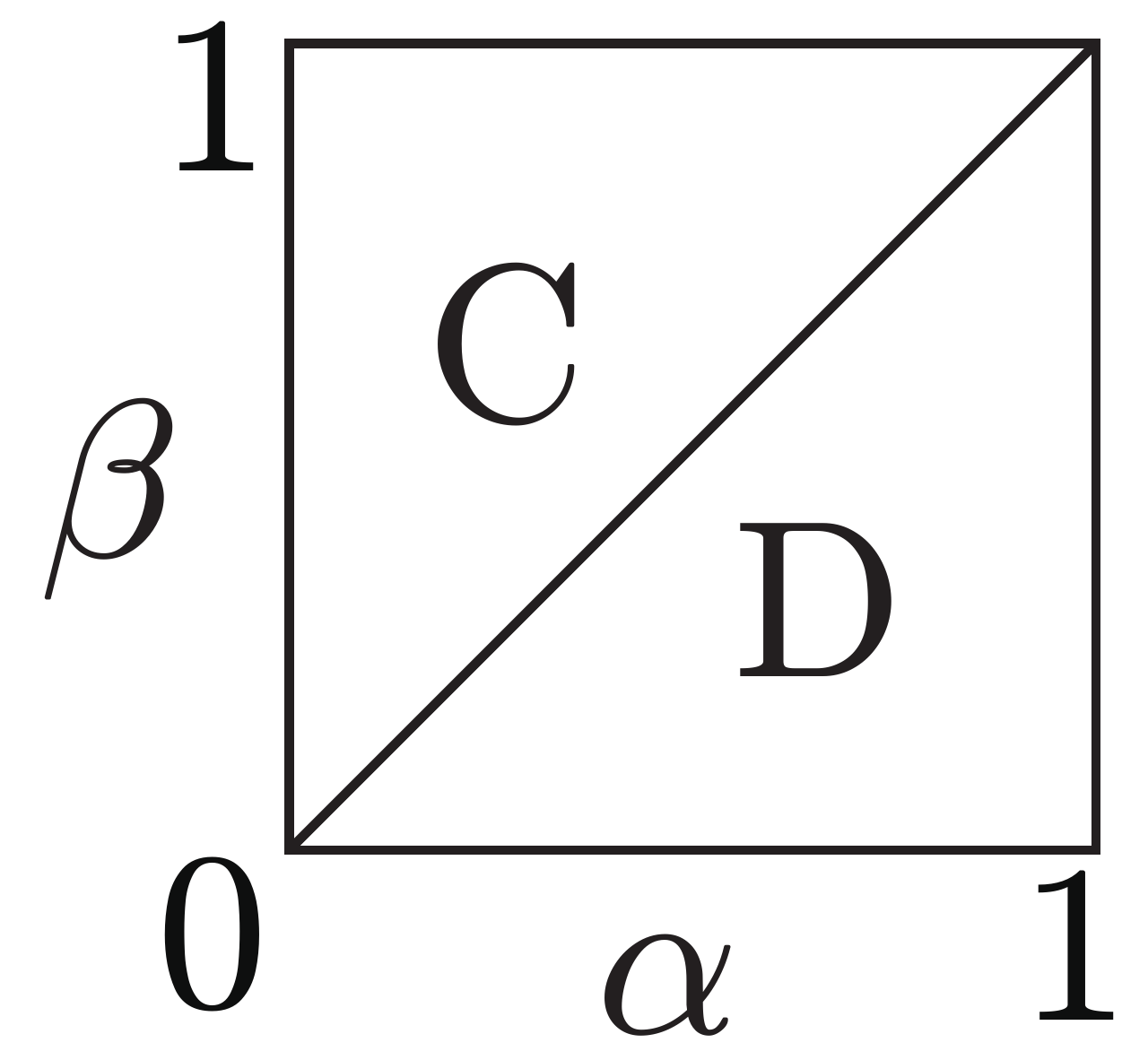}
\end{center}
\caption{Definition of M/M/1 queue (left) and its phase
diagram (right). When the arrival  probability  $\alpha$
is larger than service probability $\beta$, the queue Diverges (D).
It Converges (C) when $\alpha<\beta$. 
 \label{fig-mm1}}  
\end{figure}

The M/M/1 queueing process describes the dynamics of a single queue
with one server where arrival and service processes are Poissonian.
We usually treat it as a FIFO queue.  It is defined by the arrival
probability $\alpha$ and service probability $\beta$
\cite{bib:Medhi,bib:Saaty}.  Customers ($=$ particles) arrive with
probability $\alpha$ at the end of the queue and are serviced ($=$
removed) with probability $\beta$ at the front of the queue
(Fig.~\ref{fig-mm1}). Assuming that the particles representing
customers have unit length, the length $L_t$ of the queue at time $t$
is identical to the number of particles $N_t$.  In other words, in the
M/M/1 queueing process, the internal structure of the queue is not
considered.

For the discrete time M/M/1 queue, the probability $P(L; t)$ of having 
the length $L$ at time $t$ is governed by the master equation 
\begin{align}
P(0;t+1)  =& (1-\alpha)P(0; t)+\alpha\beta P(0;t) + (1-\alpha)\beta P(1; t)\\
\begin{split}
P(L;t+1)  =& \alpha(1-\beta)P(L-1; t) +
 [ (1 -\alpha)(1-\beta ) + \alpha\beta]  P(L;t) \\
  &  + (1 -\alpha)\beta P(L+1;t)\,.
\end{split}
\end{align}
One can easily find a stationary measure 
\begin{equation}
P_{st} (L) = \left(\frac{\alpha(1-\beta)}{(1-\alpha)\beta}\right)^L\,  .
\end{equation}
For $\alpha<\beta$, a unique stationary distribution exists, which is given
by the geometric distribution
\begin{equation}
P(L) = \frac{\beta-\alpha}{(1-\alpha)\beta}\left(\frac{\alpha(1-\beta)}{
(1-\alpha)\beta}\right)^L\,.
\end{equation}
The average length is then given by $\langle L\rangle = \sum_{L\ge 1} LP(L)$.
For $\alpha\ge \beta$, no stationary distribution exists and  
$\lim_{t\to\infty} P(L;t) =0 $ for all given $L$.
In other words, the queue diverges.
Therefore the M/M/1 queue has two phases,
according whether the queue length diverges or converges:
\begin{equation}\label{eq:mm1-L}
\lim_{t\to\infty} \langle L_t \rangle =  
\begin{cases}
\infty &\qquad \text{for }\alpha\ge \beta, \\
\frac{\alpha(1-\beta)}{\beta-\alpha}&\qquad 
\text{for }\alpha< \beta .
\end{cases}
\,
\end{equation}
The phases are separated by the {\it critical line}
  $\alpha=\beta$ (Fig.~\ref{fig-mm1}). 
%%%%%%%%%%%%%%%%%%%%%%%%%%%%%%%%%%%%%
%%%%%%%%%%%%%%%%%%%%%%%%%%%%%%%%%%%%%

Let us consider the inflow $J_\text{in}$ and outflow $J_\text{out}$ of
customers in the limit $t\to\infty$.  By definition, we always have
$J_\text{in} = \alpha$, whereas the outflow depends on the parameters.
In the convergent phase, the flows must be balanced. In fact we find
\begin{align}
    J_\text{out} =
     \beta \sum_{L\ge 1 }  P(L)  
     +  \alpha \beta    P(0) 
   = \alpha .
\end{align}
On the other hand, in the divergent phase,
the length  becomes always  larger than 0, and thus we have  
\begin{align}
    J_\text{out} = \beta . 
\end{align}

%%%%%%%%%%%%%%%%%%%%%%%%%%%%%%%%%%%%%%%%%%%%%%%%%%%%%%%%%%%%%%%%%%%%%%%%%%%%%
\section{Totally asymmetric exclusion process}
\label{sec:ASEP}

\begin{figure}
\begin{center}
  \includegraphics[width=0.55\textwidth]{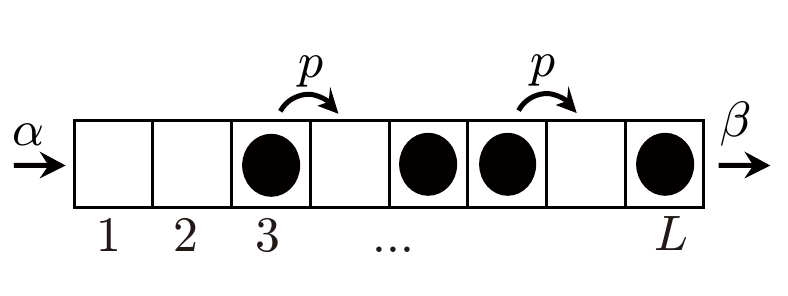}\qquad
   \includegraphics[width=0.3\textwidth]{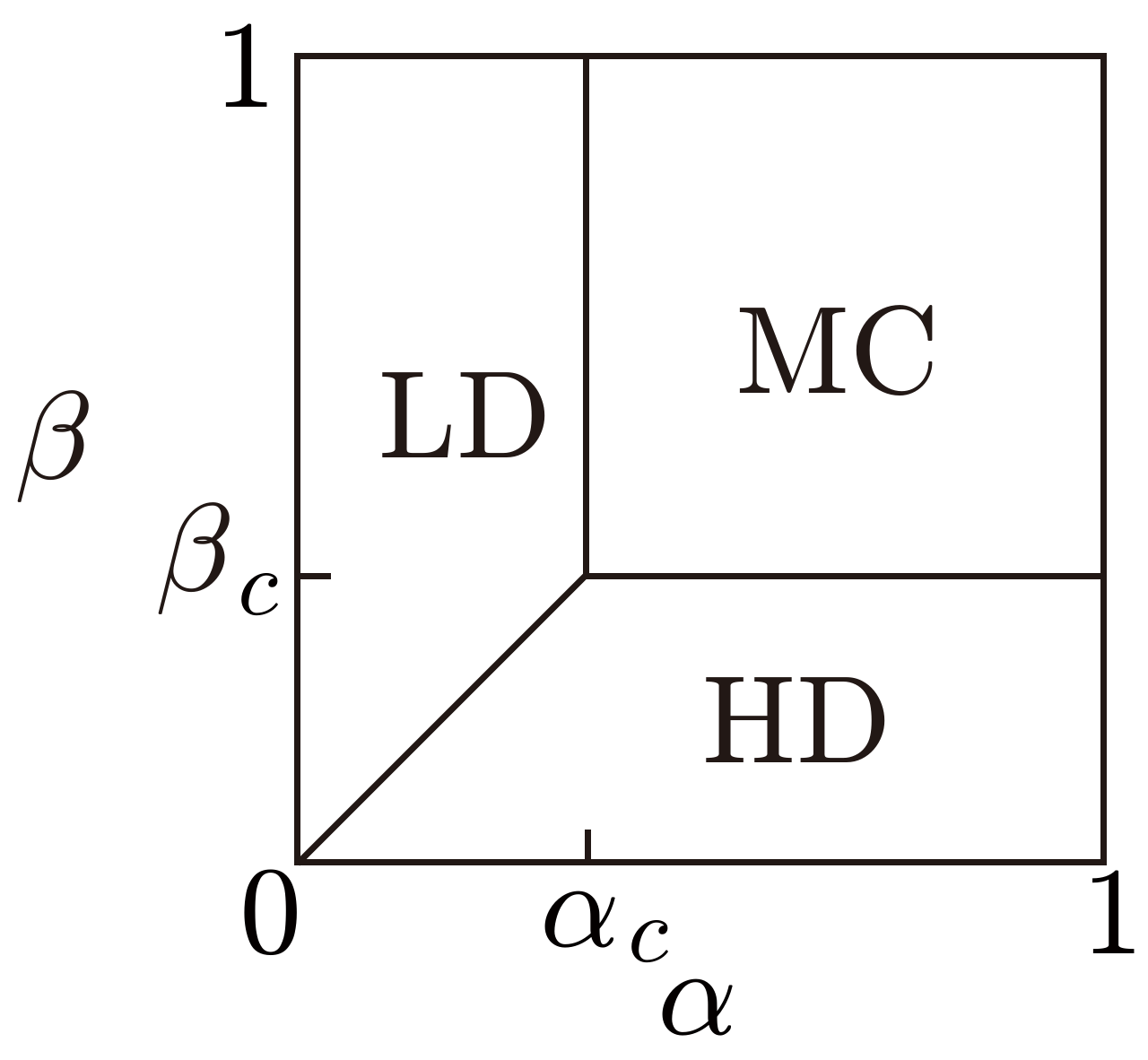}
\end{center}
\caption{\label{fig:asep}
  Definition of the TASEP (left) and its exact phase diagram for the
  parallel update (right).  The critical points are
$\alpha_c=\beta_c=1-\sqrt{1-p}$. }
\end{figure}

The Totally Asymmetric Exclusion Process (TASEP) with open boundaries
is one of the paradigmatic models of nonequilibrium physics
\cite{bib:L,bib:Schue,bib:ZS,bib:SCN,bib:KRB,bib:BE,bib:D}.  It
describes interacting (biased) random walks on a discrete lattice of
fixed length $L$, where an exclusion rule forbids occupation of a site
by more than one particle. In the TASEP illustrated in
Fig.~\ref{fig:asep}, a particle at site $j$ moves to site $j+1$ with
probability $p$ if site $j+1$ is not occupied by another particle. The
boundary sites $j=1$ and $j=L$
are coupled to particle reservoirs.  If site $1$ is
empty, a particle is inserted with probability $\alpha$.  A particle
on site $L$ is removed from the system with probability $\beta$.
%%%%%%%%%%%%%%%%%%%%%%%%%
%%%%%%%%%%%%%%%%%%%%%%%%%

Varying the boundary parameters $\alpha$ and $\beta$ (with $p $
fixed), one can distinguish three phases (Fig.~\ref{fig:asep}), namely
the Low-Density (LD), High-Density (HD) and Maximal-Current (MC)
phases.  In the LD phase the current $J= p \langle
n_j(1-n_{j+1})\rangle$ through the system depends only on the input
probability $\alpha$.  The input is less efficient than the transport
in the bulk of the system or the output and therefore dominates the
behavior of the whole system. In the HD phase the output is the least
efficient part of the system.  Therefore the current depends only on
the output probability $\beta$.  In the MC phase, input and output are
more efficient than the transport in the bulk of the system.  Here the
current has reached the maximum of the fundamental diagram, i.e.  the
relation between density and current\footnote{The fundamental
    diagram is given explicitly in Equation \eqref{eq:J-rho}.},
depending on the update rules.

The phase diagram of the TASEP was firstly derived rigorously in the
works \cite{bib:SD,bib:DEHP} for the continuous-time case. In
particular, the authors of \cite{bib:DEHP} introduced matrices to
construct the exact stationary solution. The basic idea is to make a
matrix product with replacing occupied and unoccupied sites by
matrices. This matrix product Ansatz has been widely applied to many
one-dimensional interacting particle systems \cite{bib:BE}. The matrix
product representation for the TASEP with parallel update was found in
\cite{bib:ERS} \footnote{See also ref.~\cite{bib:DeGierN} for a
  slightly different approach.}.

%%%%%%%%%%%%%%%%%%%%%%%%%%%%%%%%%%%%%%%%%%%%%%%%%%%%%%%%%%%%%%%%%%%%%%%%%%%%%
\section{Exclusive queueing process}
\label{sec:model}

\begin{figure}[t]
\vspace{-5mm}
\begin{center}
 \includegraphics[width=0.7\textwidth]{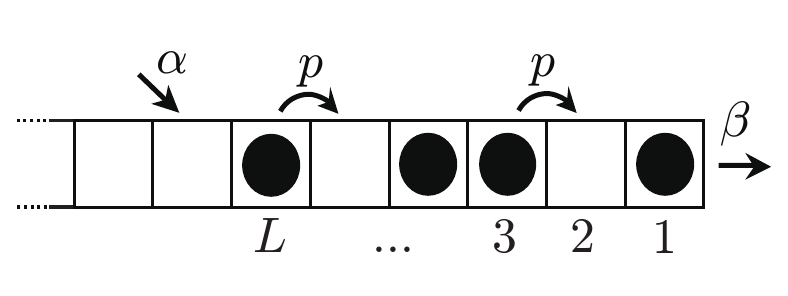}%\\
\end{center}
\vspace{-4mm}
\caption{Definition of the Exclusive Queueing Process (EQP).
 The length $L$ is defined by the position of the last
(leftmost) particle. 
\label{fig-EQP}}
\end{figure}

The Exclusive Queueing Process (EQP) is defined on a semi-infinite
chain where sites are labeled by natural numbers from right to left
(Fig.~\ref{fig-EQP}).  The dynamics of the model is defined as
follows:
\begin{description}
\item[(i)] {\it input:} A new particle is inserted with probability
  $\alpha$ on the site just behind the last particle in the queue.  If
  there is no particle in the system, a new particle is inserted
  directly to the site 1 with the same probability.
\item[(ii)] {\it hopping:} Particles behind an empty site 
 move forward with probability $p$ 
\item[(iii)] {\it output:} A particle at site 1 is serviced
  (i.e.\ removed)  with probability $\beta$.
\end{description}
For the parallel update these rules are applied simultaneously to all sites.
In the case of backward-sequential update, first
(i) and (iii) are carried out.
Then (ii) is applied sequentially to all sites starting
at site $j=1$.
The dynamics of the particle hopping  is the same as in the 
TASEP reviewed in Sec.~\ref{sec:ASEP}.
 
We define the length $L$ of the system as the position of the last
particle, and thus a new particle is inserted at site $L+1$.  Note
that this boundary condition for the left end (i) is different from
the TASEP case, whereas (iii) is the same.  Therefore the
EQP can be interpreted as a TASEP of variable length.

The EQP is formulated as a discrete-time Markov process on the state
space
\begin{equation}
S=\{\emptyset,1,10,11,100,101,110,111,1000,\dots\} , 
\end{equation}
where $0$ and $1$ correspond to unoccupied and occupied sites.
The symbol $\emptyset$ denotes the state in which there is no
particle in the queue. In the generic case, the EQP is an irreducible
and aperiodic process.

%%%%%%%%%%%%%%%%%%%%%%%%%%%%%%%%%%%%%
%%%%%%%%%%%%%%%%%%%%%%%%%%%%%%%%%%%%%

By changing the input and output probabilities $\alpha$ and $\beta$
the EQP shows boundary-induced phase transitions, 
 which are classified according to different criteria:
\begin{itemize}
\item %%%%%%%%%%%%%%%%%%%%%%%%%%%%
  Queueing classification -- convergent or divergent queue length (see
  Sec.~\ref{sec:conv-vs-div}). \\ For the parallel update this
  classification can be done by constructing exact stationary measure.
  In the Convergent (C) phase  the average system length converges to a
  finite value as $t\to\infty$.
  In the  Divergent (D) phase, the
  average length grows infinitely, being proportional to time $t$.  We
  are interested in how the phase diagram of the M/M/1 case
  (Fig.~\ref{fig-mm1}) is changed due to the excluded-volume effect.
  In Sec.~\ref{sec:deterministic}, for the special case $p=1$, we see
  that a generating function of probabilities at each time step allows
  us to rigorously compute the asymptotic behaviors \cite{bib:AS2}.
\item %%%%%%%%%%%%%%%%%%%%%%%%%%%%
  TASEP classification -- form of the outflow (see Sec.~\ref{sec:outflow}) 
  \cite{bib:AS1}. \\ 
  By definition, the inflow of particles is given by $\alpha$.  On the
  other hand, the outflow is not identical to $\beta$ because the last
  site (server) can be empty.  The form of the non-stationary flow
  $J_\text{out}$ is identical to the form for the MC  or HD phases of
  the open TASEP \footnote{This is natural since the same boundary condition
  for the right end is imposed for both the EQP and the open TASEP.}.
  In the maximal current phase, the current $J_\text{out}$  of
  particles going through the right end is independent of both
  $\alpha$ and $\beta$.
  In the high-density phase the current depends
  %only
  on $\beta$, but is independent of $\alpha$.
\item %%%%%%%%%%%%%%%%%%%%%%%%%%%%
Classification according to density profile (see Sec.~\ref{sec:subphases}).\\
The divergent phase can be divided into subphases according to the
number of plateaus in the density profile \cite{bib:AS3}.  The
rescaled profile has the form of a rarefaction wave \cite{bib:KRB}.
\end{itemize}

%%%%%%%%%%%%%%%%%%%%%%%%%%%%%%%%%%%%

\subsection{Limits and special cases}

The discrete-time EQPs have several known models as special cases or
limits.  The following diagram illustrates the relations
between the various models: 
\newcommand{\dst}{\displaystyle}
\begin{eqnarray*} \label{eq:relation-queues}
%\fl
\begin{CD}
 {\rm \fbox{Parallel EQP}} 
 @> \dst \Delta s \to 0 >> 
  {\rm \fbox{Continuous  EQP}} 
 @< \dst \Delta s \to 0 << 
  {\rm \fbox{Backward  EQP}} 
\\
%%%%%%%%%%%%%%%%%%%%%%%%%%%%%%
  @V  \dst  p=1    VV  
  @V \dst    p\to\infty   VV
  @V  \dst  p=1    VV  
 \\
%%%%%%%%%%%%%%%%%%%%%%%%%%%%%%
{\rm \fbox{ Rule 184
 %    $\begin{array}{c}
  %   {\rm Rule~184\ CA\ with} \\ {\rm stochastic\ boundaries}  
   %  \end{array}$ 
     }}
   @> \dst \Delta s\to 0 >> 
  {\rm \fbox{Continuous  M/M/1}} 
 @< \dst \Delta s \to 0 << 
  {\rm \fbox{Discrete M/M/1}}.  
\end{CD}
\end{eqnarray*}
Here we consider the {\it continuous-time limit}  $\Delta s\to 0$ 
in the master equations, which is taken as follows.
We replace $t+1$ by $t+\Delta s$, so that $\Delta s$ is the length 
of the discrete time step, and the parameters $\alpha$,
$\beta$ and $p$, and time $t$ by
$\alpha  \Delta s$, $\beta \Delta s$ and $p  \Delta s$, respectively.
In the continuous time processes, 
the parameters $\alpha$, $\beta$ and $p$  are transition rates.

The special case of the parallel EQP with $p=1$ is the rule~184
cellular automaton. Even though the hopping is deterministic, this
case is not classified into the ordinary queueing theory, since
particles  are still prohibited to jump if the preceding site is occupied.
On the other hand, the backward EQP with $p=1$ is the M/M/1 queueing
process with discrete time. The continuous-time M/M/1 queueing process
is obtained by formally taking the limit $p\to\infty$.

\subsection{Explicit probabilities}
 
\begin{figure}[t] 
\begin{center}
\includegraphics[width=0.95\textwidth]{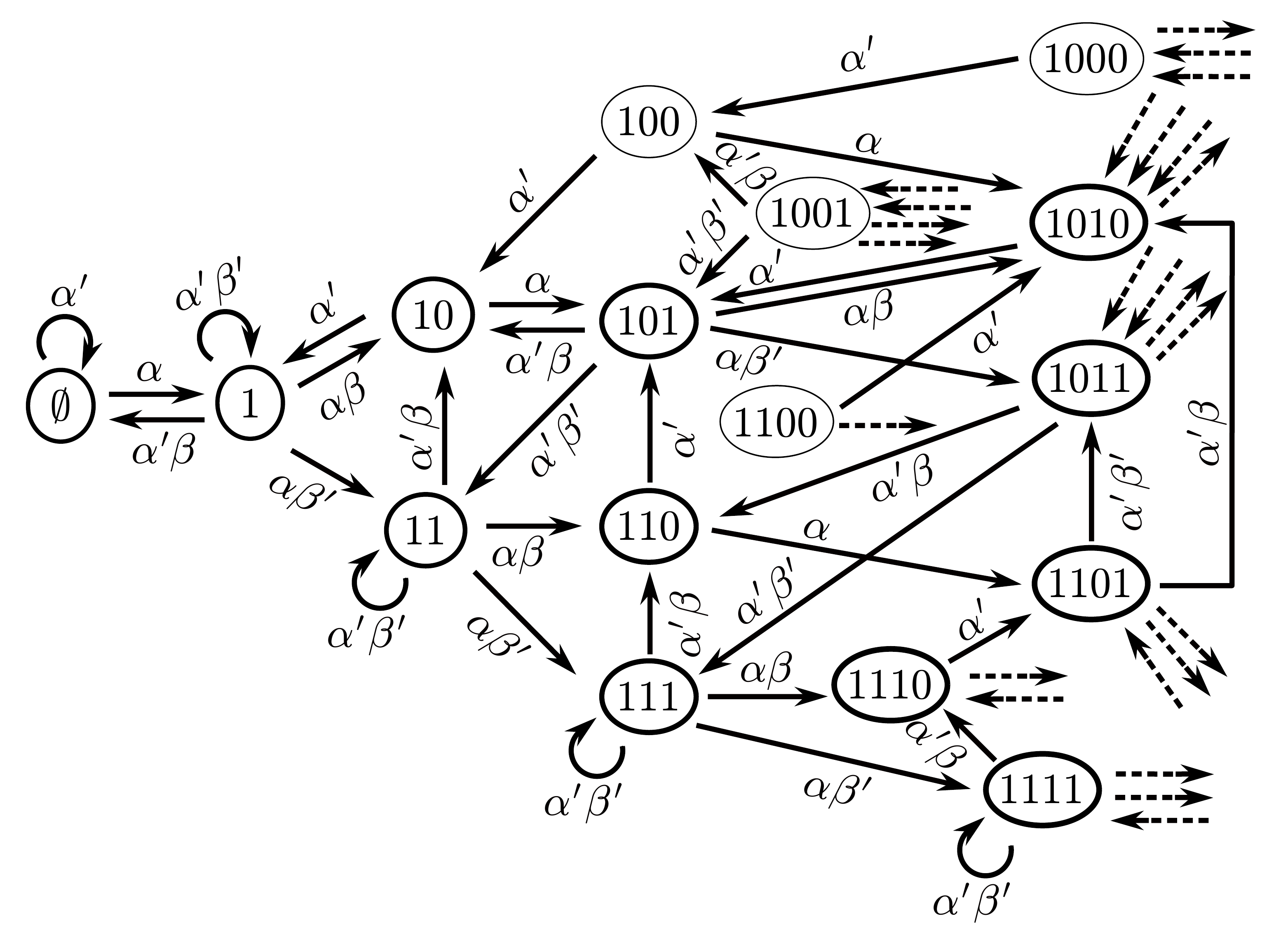}
\end{center}
\caption{Transition probabilities of a special case of the parallel EQP 
  with update with deterministic hopping ($p=1$) i.e. rule 184.  The
  bold (thin) ellipses are elements of $\tilde S $ (resp. $\tilde S
  \backslash S$).  }\label{fig:transition-probs}
\end{figure}

To close this section, we write down transition probabilities for a
few configurations with small $L$ and probabilities for a few times
steps.  We only consider the simplest case, i.e. parallel update
with $p=1$, but this is a good exercise to understand the
dynamics of the EQP.  In Fig.~\ref{fig:transition-probs}, we use
short-hand notations $ \alpha'=1-\alpha ,\beta'=1-\beta $, and the
arrows with dashed lines correspond to the transitions coming from or
going to states with length $L\ge 4$.  We notice that no arrow is
directed to configurations containing a sequence 00, which is a
consequence of the deterministic hopping. Thus this special case is
not irreducible on $S$. We restrict our consideration to the subset
$\tilde S= \{\tau\in S |\tau$ contains no 00$\}$ such that the process
is irreducible.  Let us start the process at $\emptyset$ at time
$t=0$, i.e. $ P_0(\emptyset; 0) =1$ and $ P_0(\tau; 0) =0$ for
$\tau\neq\emptyset$.  At the next time $t=1$ we have
\begin{align}
    P(\emptyset; 1) = (1-\alpha) P(\emptyset; 0) =1 -\alpha\, ,
    \qquad
    P(1; 1) = \alpha P(\emptyset; 0) = \alpha\, , 
\end{align}
and then at $t=2$ we have 
\begin{align}
    P(\emptyset; 2) = (1-\alpha ) P(\emptyset; 1) + (1-\alpha)\beta P(1; 1) 
     = (1-\alpha) (1-\alpha +\alpha\beta)\, ,
\end{align}
etc.  In principle, one can calculate all the
probabilities at any time $t$ recursively. Table~\ref{tab:probs}
provides probabilities for $t=2,3,4$. In the case of the parallel EQP
with $p=1$, a matrix product form gives probabilities for each time
step (Sec.~\ref{sec:deterministic}).

%%%%%%%%%%%%%%%%%%%%%%%%%%%%%%%%%%%%%
%%%%%%%%%%%%%%%%%%%%%%%%%%%%%%%%%%%%%

%%%%%%%%%%%%%%%%%%%%%%%%%%%%%%%%%%%%%
\begin{table}[h]
\begin{center}
\begin{tabular}{r ccccccccccc}
  $t$ & 2 & 3 & 4 \\ \hline $\emptyset $ & $ \alpha'(\alpha' +
  \alpha\beta)$ & $ \alpha'^2 ( 2\alpha\beta + \alpha' $ & $ \alpha'^2
  ( 2\alpha^2\beta^2 + 3\alpha\beta\alpha' + \alpha'^2$ \\ & & $ +
  \alpha\beta\beta' ) $ & $ + 2\alpha\beta\alpha'\beta'
  +\alpha\beta\alpha'\beta'^2) $ \\ \hline 1 & $\alpha \alpha'
  (1+\beta') $ &$\alpha\alpha'(2\alpha\beta+\alpha' $ & $
  \alpha^2\alpha'\beta (3\alpha\beta+\alpha' +5\alpha\beta\beta' $ \\ 
  & & $ +\alpha'\beta'+\alpha'\beta'^2) $ & $
  +\alpha'\beta'+\alpha'\beta'^2+\alpha'\beta'^3) $ \\ \hline 10 &
  $\alpha^2\beta $ & $\alpha^2\alpha'\beta (1+2\beta')$ & $
  \alpha^2\alpha'\beta (3\alpha\beta+\alpha'+
  2\alpha'\beta'+3\alpha'\beta'^2) $ \\ 11 & $\alpha^2\beta'$ &
  $\alpha^2\alpha'\beta'(1+2\beta')$ & $ \alpha^2\alpha'\beta
  (3\alpha\beta+\alpha'+ 2\alpha'\beta'+3\alpha'\beta'^2) $ \\ \hline
  101 & 0 & $\alpha^3\beta $ & $\alpha^3\alpha'\beta(1+3\beta')$ \\ 
  110 & 0 & $\alpha^3\beta\beta' $ &
  $\alpha^3\alpha'\beta\beta'(1+3\beta')$ \\ 111 & 0 &
  $\alpha^3\beta'^2$ & $\alpha^3\alpha'\beta'^2(1+3\beta')$ \\ \hline
  1010 & 0 & 0 & $\alpha^4\beta^2 $ \\ 1011 & 0 & 0 &
  $\alpha^4\beta\beta' $ \\ 1101 & 0 & 0 & $\alpha^4\beta\beta' $ \\ 
  1110 & 0 & 0 & $\alpha^4\beta\beta'^2 $ \\ 1111 & 0 & 0 &
  $\alpha^4\beta'^3$
\end{tabular}
\end{center}
\label{tab:probs}
\caption{Probabilities of finding configurations at first few time 
  steps for the rule 184 case.}
\end{table} 
%%%%%%%%%%%%%%%%%%%%%%%%%%%%%%%%%%%%%

%%%%%%%%%%%%%%%%%%%%%%%%%%%%%%%%%%%%%%%%%%%%%%%%%%%%%%%%%%%%%%%%%%%%%%%%%%%%%

\section{Convergent and divergent phases}\label{sec:conv-vs-div}

For the parallel EQP the stationary measure has the following matrix
product form, which provides the grand canonical ensemble of the
parallel-update TASEP with open boundaries, as explained in
Sec.~\ref{sec:ASEP}, with $\alpha=1$:
\begin{eqnarray} 
\label{eq:MP}
 P_{st}(\emptyset)&=& 1\,, \label{arrmpf1}\\
 P_{st}(1\tau_{L-1}\dots\tau_1)
 &=& \left(\frac{\alpha}{p(1-\alpha)} \right)^L
   \langle W | DX_{\tau_{L-1}}\cdots X_{\tau_1} |V\rangle\,.\label{arrmpf2}
\end{eqnarray}
where $\frac{\alpha}{p(1-\alpha)} $ plays a role of a fugacity.  The
matrices $X_1=D$ and $X_0=E$, the row vector $\langle W|$ and the
column vector $|V\rangle$ should satisfy quartic algebraic relations
which are identical to those for the parallel-update TASEP with open
boundaries and $\alpha=1$ \cite{bib:ERS}.  The matrix product form
\eqref{eq:MP} allows to use some exact results obtained for the
parallel TASEP.  The representations of the matrices and vectors are,
in general, infinite-dimensional \cite{bib:ERS}.
For the special case $p=1$, the matrices and vectors have two dimensional
representation:
\begin{align}\label{eq:two-by-two}
  D= \left(\begin{array}{cc} 1/\beta-1 & 0 \\
      1/\sqrt{\beta} & 0  \end{array}\right),\ 
 E= \left(\begin{array}{cc} 0  &  1/\sqrt{\beta} \\0 & 0 \end{array}\right), \ 
 \langle W| = \left( 1 \  \sqrt{\beta}  \right) , \quad
  |V\rangle  = \left(\begin{array}{c} 1 \\  \sqrt{\beta}   \end{array}\right) .
\end{align}
On the other hand, by taking continuous-time limit we obtain the
matrix product stationary measure for the continuous-time EQP
\cite{bib:A}, whose algebra corresponds to the continuous-time TASEP
with open boundaries \cite{bib:DEHP}.

As far as we know, a physical interpretation of the grand-canonical
ensemble to a process with varying system length was firstly shown in
\cite{bib:Heil}.  A similar construction i.e. a matrix product with
fugacity, is also possible for a simple model of microtubule growth
\cite{bib:MRF,bib:ALS,bib:Lueck}.  However, this is not always true
for all TASEPs with varying length.  For example the EQP with the
backward update, a matrix product form has not been found, although
the open TASEP with backward update has a matrix product stationary
state \cite{bib:Evans,bib:RSS,bib:RSSS}.  In the recent work
\cite{bib:dGF}, a queueing process with two types of customers was
introduced, which is called Prioritizing Exclusion Process (PEP).
High priority customers can overtake low priority customers.  This is
another variant of the TASEP with varying system length, by regarding
high- and low-priority customers as particles and empty sites,
respectively. However, a matrix product stationary measure for the PEP
has not been found up to now.

For the parallel EQP,  the series 
\begin{equation}
Z = \sum_{\tau\in S} P_{st}(\tau)
 = 1+ \sum_{L\ge 1 }  \left(\frac{\alpha}{p(1-\alpha)} \right)^L
   \langle W | D (D+E)^{L-1} |V\rangle 
\end{equation}
converges, when the condition 
\begin{equation}
\begin{cases} 
 \alpha\le\alpha_c=\frac{1-\sqrt{1-p}}{2} &
     \text{for } \beta> \beta_c = 1-\sqrt{1-p},\\
 \alpha<\alpha_c=\frac{\beta(p-\beta)}{p-\beta^2} & \text{for } 
  \beta\le \beta_c  
\end{cases}
\label{eq:alpha<alpha_c-parallel}
\end{equation}
is satisfied \cite{bib:AY,bib:ERS}.  The existence of the stationary
distribution $ \frac{1}{Z}P_{st}(\tau) $ guaranties that we will
approach to it, starting from any initial state.  In the region
$\alpha<\alpha_c $ (convergent phase), the average system length
$\langle L_t \rangle $ and the average number of particles $\langle
N_t \rangle$ converge to
\begin{align}\label{eq:LN-staitonary} 
   \langle L_\infty \rangle  =  \frac{\alpha p(R-p+2(1-\alpha))}{R(R-p
+2(1-\alpha)\beta)},\quad  
 \langle N_\infty \rangle =
    \frac{\alpha (1-\alpha)(p-2\alpha p+R)}{R(R-p+2(1-\alpha)\beta)} 
\end{align}
with $R=\sqrt{p(p-4\alpha(1-\alpha))}$. 
Oppositely, for $\alpha>\alpha_c$ (divergent phase)
and for $\alpha=\alpha_c$ (critical line),
 $\langle L_t \rangle $ and  $\langle N_t  \rangle$ diverge.
On the straight-line part of the critical line
$\alpha=\alpha_c$ ($\beta>1-\sqrt{1-p} $), there are distributions of
the system length and the number of particles, but their averages diverge.
When $p=1$, the condition (\ref{eq:alpha<alpha_c-parallel}) simplifies to\footnote{
The  eigenvalues of $D+E$ for \eqref{eq:two-by-two} is 
$\{ -1,  1/\beta \}$, and the critical value can be derived by
$  \frac{\alpha_c }{1 - \alpha_c}\frac 1 \beta =1 $.
}
\begin{equation}
 \alpha < \alpha_c= \frac{\beta}{1+\beta}. 
\end{equation}

As we mentioned, we could not find an exact stationary measure for the
backward case. Thus the determination of the phase diagram has to be
done by Monte Carlo simulations. The region where the average system
length converges is expected to be
\begin{eqnarray}
\begin{cases} 
\label{eq:alpha<alpha_c-back}
     \alpha \le   \alpha_c  =  \frac{(1-\sqrt{1-p})^2 }{p}
       & (\beta_c < \beta < 1)  ,\\
    \alpha < \alpha_c  =  \frac{\beta(p-\beta)}{p(1-\beta)}
       & (0<\beta\le \beta_c ) .
   \end{cases} 
\end{eqnarray}
For the backward case, explicit forms of the average values like
\eqref{eq:LN-staitonary} are unknown except for the special case
$p=1$.

%%%%%%%%%%%%%%%%%%%%%%%%%%%%%%%%%%%
%\subsection{Dynamics for deterministic hopping}
 \section{Dynamics for deterministic hopping}\label{sec:deterministic}

In the case of parallel dynamics, an exact time-dependent solution is 
also known for deterministic hopping $p=1$ in the bulk \cite{bib:AS2}.  

For the deterministic hopping case, an exact expression of the
``dynamical state'' is possible; starting from the empty chain
$\emptyset$ at $t=0$, the probability $ P(\tau;t) $ of finding a state
$ \tau $ at time $t$ can be written as
\begin{align}
  P(\emptyset;t) &= \oint \frac{dz}{2\pi iz^{t+1}} \frac{1-\Lambda}{1-z},
\\
  P(\tau_L\cdots\tau_1;t) 
  &=   \langle W| X_{\tau_L} \cdots X_{\tau_1}  |V\rangle  
  \beta^L \oint \frac{dz}{2\pi iz^{t+1}} 
  \frac{1-\Lambda }{1-z}  \Lambda^L
\end{align}
with the same matrices and vectors as for the stationary case
\eqref{eq:two-by-two}, and the fugacity $ \Lambda =\frac{1- \alpha'
  \beta' z - \sqrt{ ( 1-\alpha' \beta' z )^2 - 4 \alpha \alpha' \beta
    z^2 } }{2 \alpha'\beta z} $.  The contour integral picks up the
coefficient of $ z^t $ in the power series of
$\frac{1-\Lambda}{1-z}\Lambda^L$. This simple form is due to the
simplification of the master equation in the special case $p=1$, see
\cite{bib:AS2} for details.

The probability of finding the length $L$ at time $t$ is given as
$\oint \frac{dz}{2\pi iz^{t+1}} \frac{1-\Lambda}{1-z} \Lambda^L$,
since $\langle W| D (D+E)^{L-1} |V\rangle \beta^L = 1 $.  Thus the
average length of the system at time $t$ and its asymptotic behaviors
($t\to\infty$) are given by
\begin{align}
\langle L_t\rangle 
=\oint  \frac{dz}{2\pi iz^{t+1}} \frac{\Lambda}{  (1-z)(1-\Lambda)  }   
 \simeq
 \left\{\begin{array}{ll}
   \frac{\alpha}{\beta-\alpha-\alpha\beta}
    &  \quad (\alpha<\frac{\beta}{1+\beta} ),\\
  2\sqrt{  \frac{\beta t}{\pi (1+\beta )}  }
    &  \quad (\alpha=\beta ),\\
  (\alpha-\beta + \alpha \beta)  t 
  &  \quad (\alpha>\frac{\beta}{1+\beta} ) .
 \end{array}\right.
\label{eq:Lsim}
\end{align}
The probability of finding  $N$ particles  at time $t$ is given as 
$\oint \frac{dz}{2\pi iz^{t+1}} \frac{1-\Lambda }{(1-z)}$  (for $N=0$) and 
$\oint \frac{dz}{2\pi iz^{t+1}} 
\frac{ \Lambda(1-\Lambda)(1+\beta\Lambda)}{(1-z)}
  [\Lambda(1-\beta+\beta\Lambda)]^{N-1}$  (for $N\ge 1$).
The average number of particles at time $t$  and its asymptotic behavior  
($t\to\infty$) can be also  calculated as 
\begin{align}
\begin{split}
 \langle N_t\rangle 
  = \oint  \frac{dz}{2\pi iz^{t+1}} 
\frac{\Lambda}{(1-z)(1-\Lambda)(1+\beta\Lambda)}
\end{split}
  \simeq  
 \left\{\begin{array}{ll}
   \frac{\alpha(1-\alpha)}{\beta-\alpha-\alpha\beta} 
   & \ \  (\alpha<\frac{\beta}{1+\beta}  ),\\
   2\sqrt{ \frac{\beta t}{\pi(1+\beta)^3}  } 
     & \ \  (\alpha=\frac{\beta}{1+\beta} ),\\
   \frac{\alpha-\beta +\alpha\beta}{1+\beta} t
    & \ \ (\alpha>\frac{\beta}{1+\beta} )  .
 \end{array}\right.
\end{align}
The asymptotic behaviors on the critical line are diffusive (i.e.
$\sim t^{1/2}$) as the symmetric random walk. For general $p$,
however, more complicated behavior is observed (see
Sec.~\ref{sec:critical}).

%%%%%%%%%%%%%%%%%%%%%%%%%%%%%%%%%%%%%%%%%%%%%%%%%%%%%%%%%%%%%%%%%%%%%%%%%%%%%

\section{The outflow}\label{sec:outflow}

Let us consider the non-stationary properties of the EQP in order to
derive the phase diagram  based on physical arguments.  This
heuristic understanding of the phase diagram is similar to the open TASEP case,
where a domain wall between a low- and high-density  regions
($\rho_\text{left}$ and $\rho_\text{right}$, respectively) moves 
with velocity   \cite{RefKSKS}   
\begin{equation}\label{eq:v_s}
  v_s= \frac{J(\rho_\text{left}) -J(\rho_\text{right}) }
{\rho_\text{left} -\rho_\text{right} }  .
\end{equation}
Here the fundamental diagram  $J=J(\rho)$ depends on the update 
rule \cite{bib:RSS,bib:RSSS}.

For the EQP, from the particle conservation law, we have 
\begin{align}
  \langle N_t \rangle = (  J_\text{in} - J_\text{out}  ) t
   +   \langle N_0 \rangle 
\end{align}
Here we assume the outflow $ J_\text{out} $ is a constant. 
On the other hand, 
the inflow $J_{\rm in}$ is always $\alpha$, which is due to the fact
that the site where particles enter is by definition never blocked.

From Monte Carlo simulations, we find the outflow $ J_\text{out}$ as
\begin{equation}\label{eq:Jout-para}
   J_\text{out}  = 
 \begin{cases}
\frac{\beta(p-\beta)}{(p-\beta^2) }& \quad (\beta\le \beta_c ),  \\ 
 \frac{(1-\sqrt{1-p})}{2} & \quad  (\beta> \beta_c ) , 
\end{cases}
  \end{equation}
for the parallel case and  
 \begin{equation}\label{eq:Jout-back}
\quad 
    J_\text{out}  = 
\begin{cases} 
    \frac{\beta(p-\beta)}{p(1-\beta)}  & \ (\beta\le \beta_c ),\\
    \frac{(1-\sqrt{1-p})^2 }{p}        & \  (\beta_c < \beta).
   \end{cases} 
\end{equation}
for the backward case.  According to the TASEP explained in
Sec.~\ref{sec:model} these phases might be called High-Density (HD)
phase for $\beta\le \beta_c$ and Maximal-Current (MC) phase for
$\beta_c < \beta$.  Note that the form of the outflow is identical to
the critical value $\alpha_c$ as given in
Equations~\eqref{eq:alpha<alpha_c-parallel} and
\eqref{eq:alpha<alpha_c-back}. The phase diagram is now understood as
follows: when $J_\text{in} < J_\text{out}$ ($J_\text{in} >
J_\text{out}$), the number of particles increases (resp.\ decreases).
The system length also increases (resp.\ decreases) according to
$J_\text{in} < J_\text{out}$ ($J_\text{in} > J_\text{out}$), if we
assume the ``density'' $ \langle N_t \rangle / \langle L_t \rangle $
is a constant.  We remark that the density profile is not always
globally constant, which we will review in the next Section.  We also
remark that the formula \eqref{eq:v_s} is {\it not} satisfied by the
``shock'' (i.e. the left end of the density profile) in the EQP.
After the ``shock'' reaches the vicinity of the server, the forms
\eqref{eq:Jout-para} and \eqref{eq:Jout-back} are no longer valid, and
the outflow becomes $\alpha$. 
This means the convergence to the stationary distribution.

 So far, based on the queueing and TASEP classifications, we
have divided the parameter space into 4 regions, the
MC-C, MC-D, HD-C and HD-D phases.

%%%%%%%%%%%%%%%%%%%%%%%%%%%%%%%%%%%%%%%%%%%%%%%%%%%%%%%%%%%%%%%%%%%%%%%%%%%%%
\section{Subphases of the divergent phase}\label{sec:subphases}

Let us consider the situation that the input probability $\alpha $ is
much larger than the output probability $\beta$ (e.g. $\alpha =1 $).
In this case, new particles are always injected to the system, so the
density  $\rho_\text{left}$  near the left end is expected to be 1.  On the
other hand, the density near the right end is expected to be
\begin{eqnarray} 
\rho_\text{right} \simeq \begin{cases}
 \frac{p-\beta}{p-\beta^2}   & \quad (\beta\le\beta_c),\\
 \frac 1 2                & \quad (\beta>\beta_c ) 
\end{cases}
\end{eqnarray}
 for parallel update and  
\begin{eqnarray}
 \rho_\text{right} \simeq  \begin{cases}
 \frac{p-\beta}{p(1-\beta)}  & \quad (\beta\le\beta_c) , \\
  \frac{1-\sqrt{1-p}}{p}  & \quad (\beta>\beta_c )
\end{cases}
\end{eqnarray}
 for backward update. 
In this section, we review the global density profile that can obtained
by ``cutting'' a rarefaction wave, and we  will see that the density 1
near the left end is not always realized.
Then we further divide the divergent phases into subphases based on
the form of the density profile.

\begin{figure} 
\begin{center}
 \includegraphics[width=0.42\textwidth]{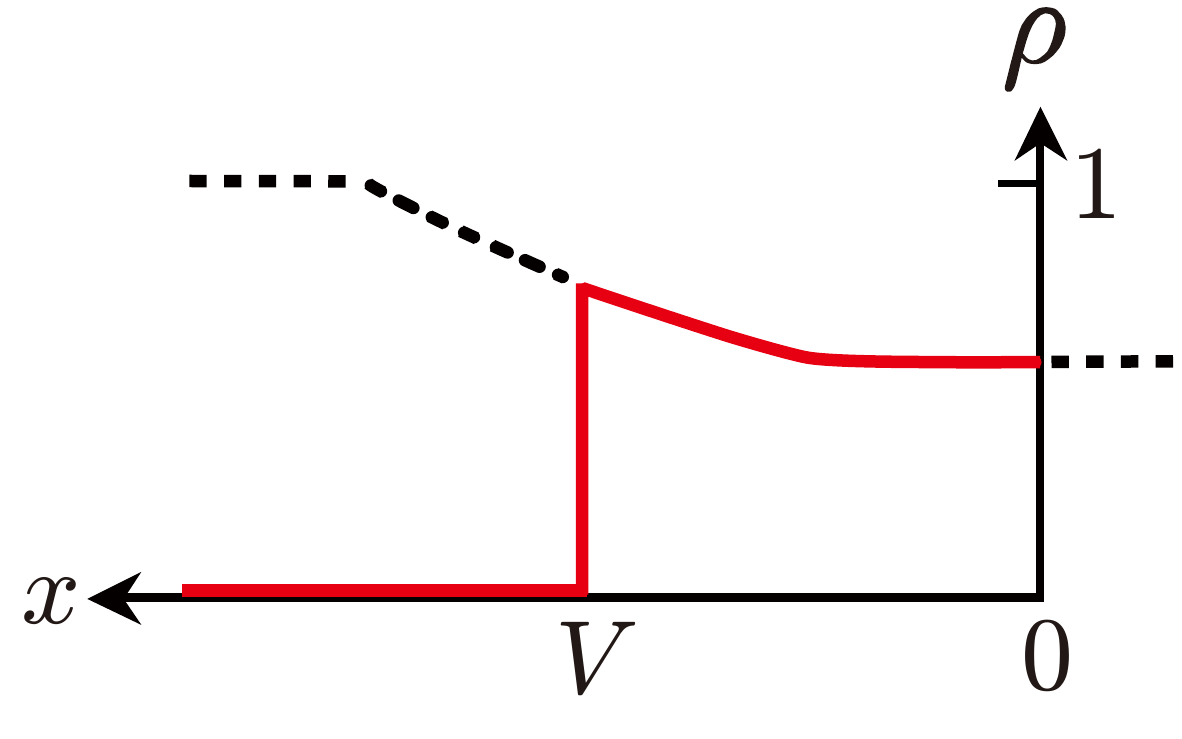}
\quad \quad 
 \includegraphics[width=0.42\textwidth]{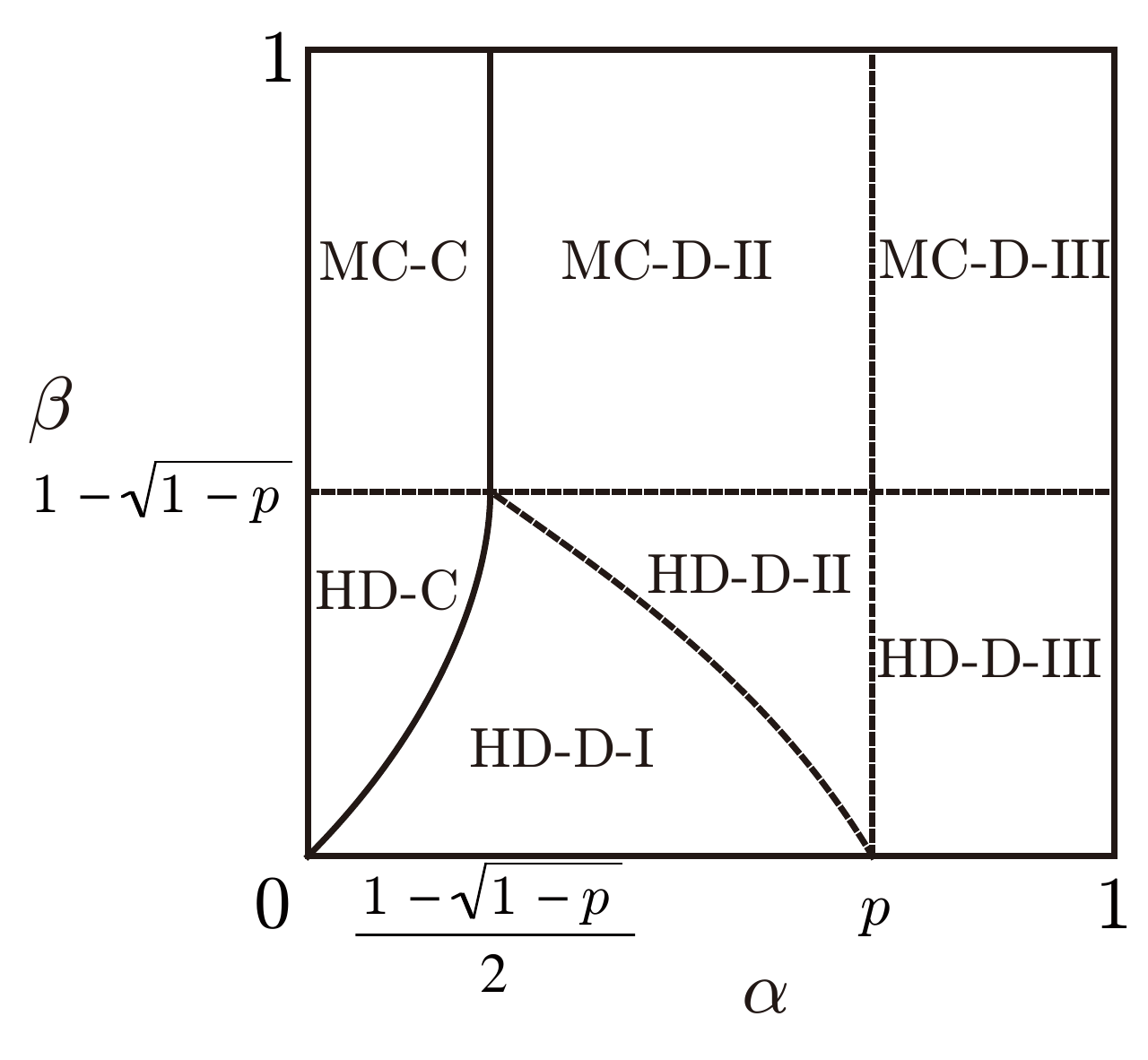}
 \includegraphics[width=0.42\textwidth]{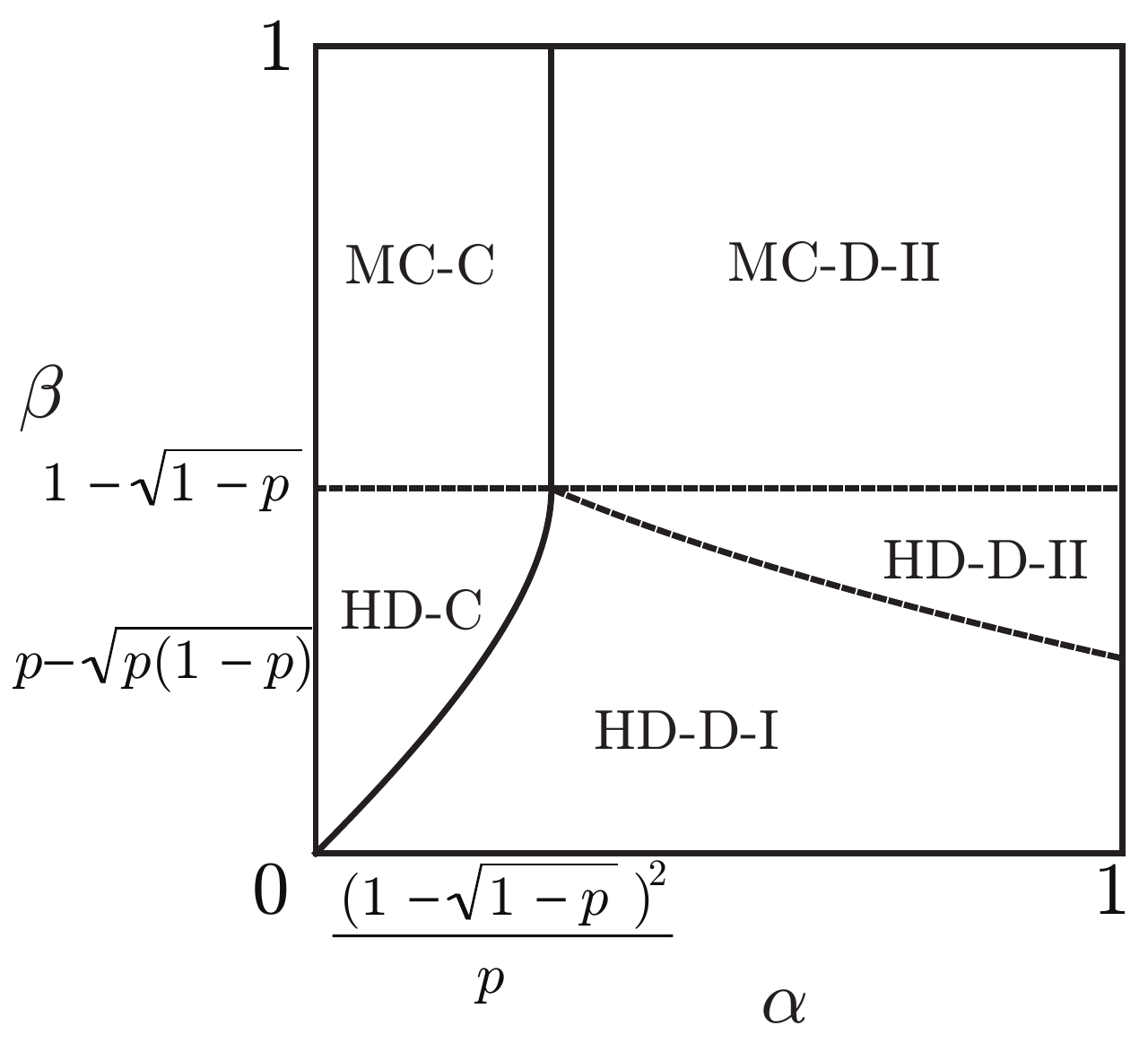}
\qquad
 \includegraphics[width=0.42\textwidth]{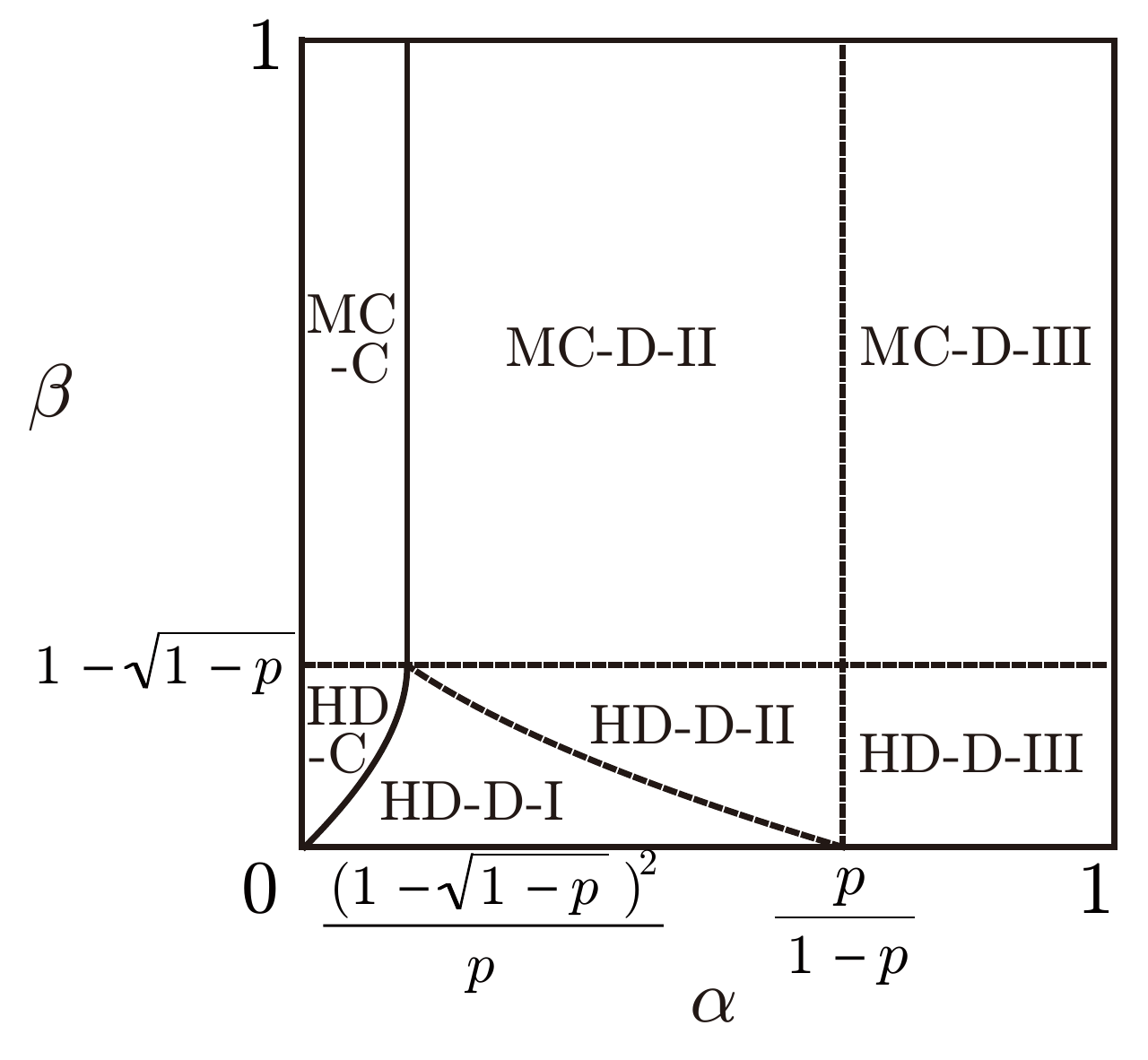}
\end{center}
\caption{A schematic picture for the density profile
  in the divergent phase (top-left), where $x$ is the rescaled
  position $j/t$, and the phase diagrams of the EQP with parallel
  (top-right) and backward-sequential dynamics for $p<1/2$
  (bottom-left) and $p>1/2$ (bottom-right).\label{fig-phases}
  According to the injection probability (rate) $\alpha$, the
  rarefaction wave is ``cut'' by the leftmost particle ($x=V$) and the
  server ($x=0$). The end of the queue can be in three different
  regimes, (plateau at density $\rho<1$, regime of increasing density
  or plateau at density $\rho=1$), which defines the regions I, II,
  III., respectively.  The density profile shown here belongs to II.
  \label{fig-profile} }
\end{figure}

In the  TASEP (typically on $\mathbb Z$), 
a rarefaction wave is derived by a hydrodynamic approach \cite{bib:KRB}:
The rescaled density profile $\rho(x=j/t)$ (see Fig.~\ref{fig-profile}) 
\begin{equation}
\label{eq:rho(x)}
  \rho (x) \simeq
  \begin{cases}
   \rho_{\rm right} & \qquad(x<f(\rho_{\rm right})  ),  \\
   f^{-1}(x)        & \qquad(f(\rho_{\rm left})>x>f(\rho_{\rm right})), \\
   \rho_{\rm left}  & \qquad(x>f(\rho_{\rm left})  )  
  \end{cases} 
\end{equation}
with $f(\rho) = -\frac{dJ}{d \rho}$ 
does not change the shape (i.e. is ``time invariant'') 
for $  \rho_{\rm left} > \rho_{\rm right}$.
The fundamental diagram is given as 
\begin{equation}
\label{eq:J-rho}
J (\rho)=\begin{cases}
\frac{1-\sqrt{1-4p\rho(1-\rho)}}{2} &\qquad (\text{parallel}),\\
\frac{p\rho(1-\rho) }{1-p\rho }     &\qquad (\text{backward}).
\end{cases}
\end{equation}
Thus we find the curved part of the profile, respectively, as 
\begin{equation}
f^{-1} (x) = \begin{cases}
\frac{1}{2} + \frac{x}{2} \sqrt{\frac{1-p}{p(p-x^2)}} & \qquad 
\text{(parallel}),\\
\frac{1}{p} - \frac{1}{p}\sqrt{ \frac{ 1-p}{ 1+x} } & \qquad 
\text{(backward}).
\end{cases}
 \end{equation}

%%%%%%%%%%%%%%%%%%%%%%%%%%%%%%%%%%%%%
%%%%%%%%%%%%%%%%%%%%%%%%%%%%%%%%%%%%%

We assume that in the divergent phase of the EQP the global density
profile $\rho (x) $  can be written as \eqref{eq:rho(x)} for $ 0<x<V$,
where the velocity\footnote{ Note that $x=j/t$ has the dimension
    of a velocity. } $V$  (i.e. $V=\langle  L_t \rangle /t $)
    is determined by the particle number
  conservation: $\int_0^V \rho (x)dx = \alpha -J_\text{out}$.  From
  this assumption, which is supported by simulations, we find
\begin{equation}\label{eq:velo-L-para}
V =  \begin{cases}
         \alpha\frac{p-\beta^2}{p-\beta} - \beta
        & \left(\text{I:  }            \frac{\beta(p-\beta)}{p-\beta^2}
   <\alpha \le \frac{(p-\beta)^2}{p-2p\beta+\beta^2} 
        \right), \\
       2p\alpha-p+2\sqrt{p\alpha(1-p)(1-\alpha)}
        & \left(\text{II:  }
        \max\left(\frac{(p-\beta)^2}{p-2p\beta+\beta^2}, 
          \frac{1-\sqrt{1-p}}{2}  \right)  < \alpha \le p \right), \\
       \alpha & \left(
       \text{III:  }  \quad p < \alpha \le 1       \right) , 
  \end{cases}   
\end{equation}
for the parallel EQP, and 
\begin{equation}
  V =
   \begin{cases}
  \frac{p(1-\beta)}{p-\beta}\alpha-\beta
        & \left(\text{I: } 
             \frac{\beta(p-\beta)}{p(1-\beta)}
   <\alpha \le \frac{(p-\beta)^2}{p(1-p)} \right), \\
       2\sqrt{p (1-p)\alpha -p(1-\alpha ) }
        & \left(\text{II: }
        \max \left(  \frac{(p-\beta)^2}{p(1-p)} ,
 \frac{(1-\sqrt{1-p})^2  }{p}  \right)  < \alpha \le \frac{p}{1-p} 
        \right), \\
       \alpha & \left(
       \text{III: } \frac{p}{1-p} < \alpha \le 1  \right),
  \end{cases} 
\label{eq:velocitybackward}
\end{equation}
for the backward EQP.

To summarize, we have found up to 5 subphases in the divergent case,
according to the classification based on the forms of the outflow $J_\text{out}$
and the velocity  $V$, as shown Fig.~\ref{fig-profile}.  The shape of the global
density profile changes depending on the parameters.

%%%%%%%%%%%%%%%%%%%%%%%%%%%%%%%%%%%%%%%%%%%%%%%%%%%%%%%%%%%%%%%%%%%%%%%%%%%%%
\section{Critical line: Non-universal behavior}
\label{sec:critical}

In the divergent phase the average length $\langle L_t\rangle$ and
the average number of particles $\langle N_t\rangle$ diverge linearly
in time. On the phase transition line separating the convergent
and divergent phases the growth is slower than linear, i.e. 
 \begin{equation}
\langle X_t \rangle 
= O ({ t^{\gamma_X}} ) 
  \qquad
   (X=L,N)  ,
 \end{equation}
where the critical exponents $\gamma_X$ are smaller than 1.

\begin{figure}
\begin{center}
 \includegraphics[width=0.42\textwidth]{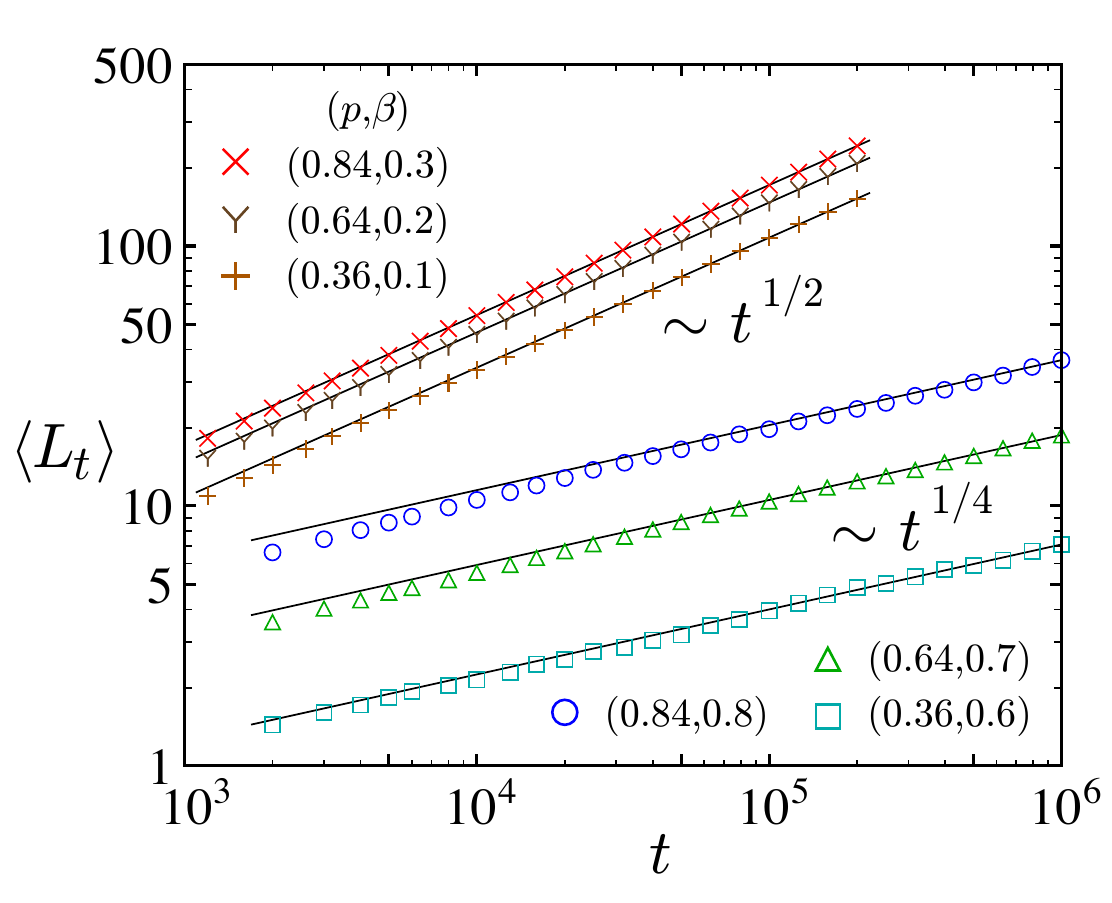}
\qquad
 \includegraphics[width=0.42\textwidth]{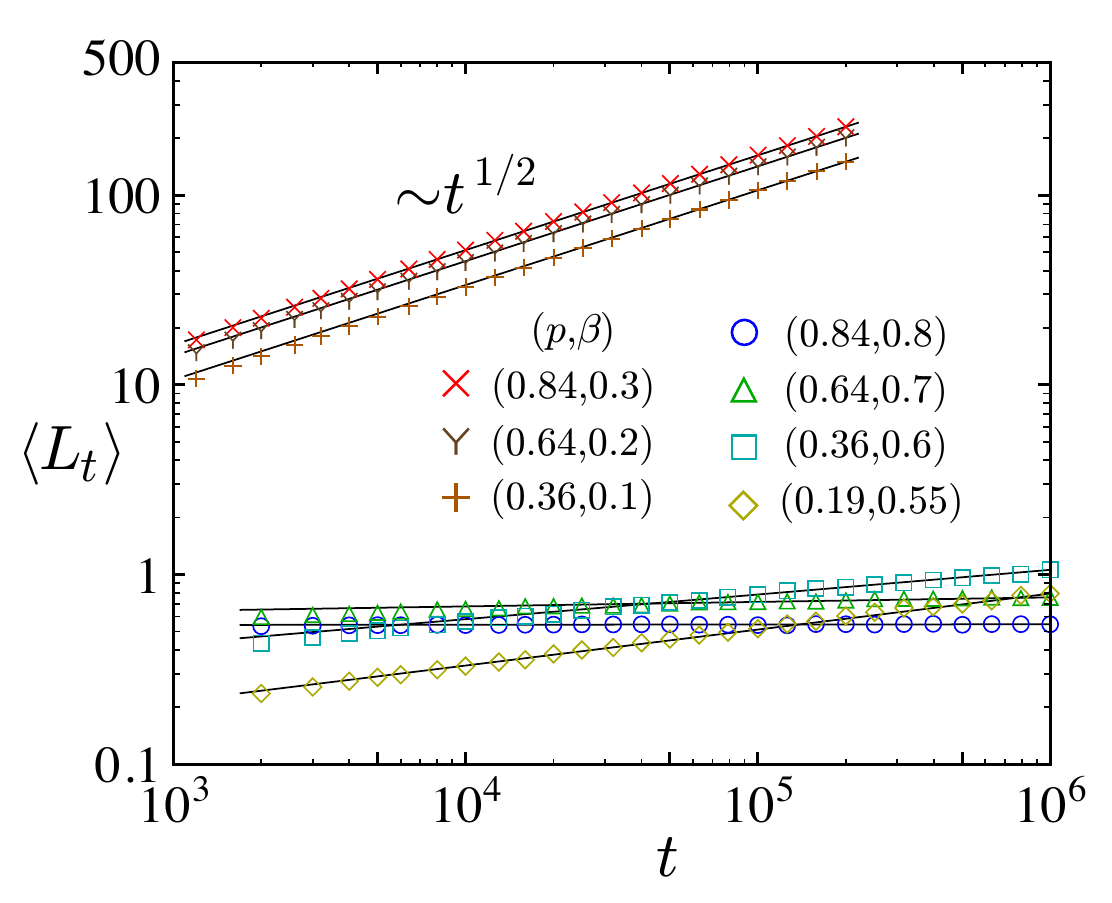}
\end{center}
\vspace{-5mm}
\caption{Time-dependence of average system length $\langle L_t\rangle$ 
 on the critical line  for 
parallel dynamics (left) and backward dynamics (right).
\label{fig-length}}
\end{figure}

\begin{figure}
\begin{center}
 \includegraphics[width=0.42\textwidth]{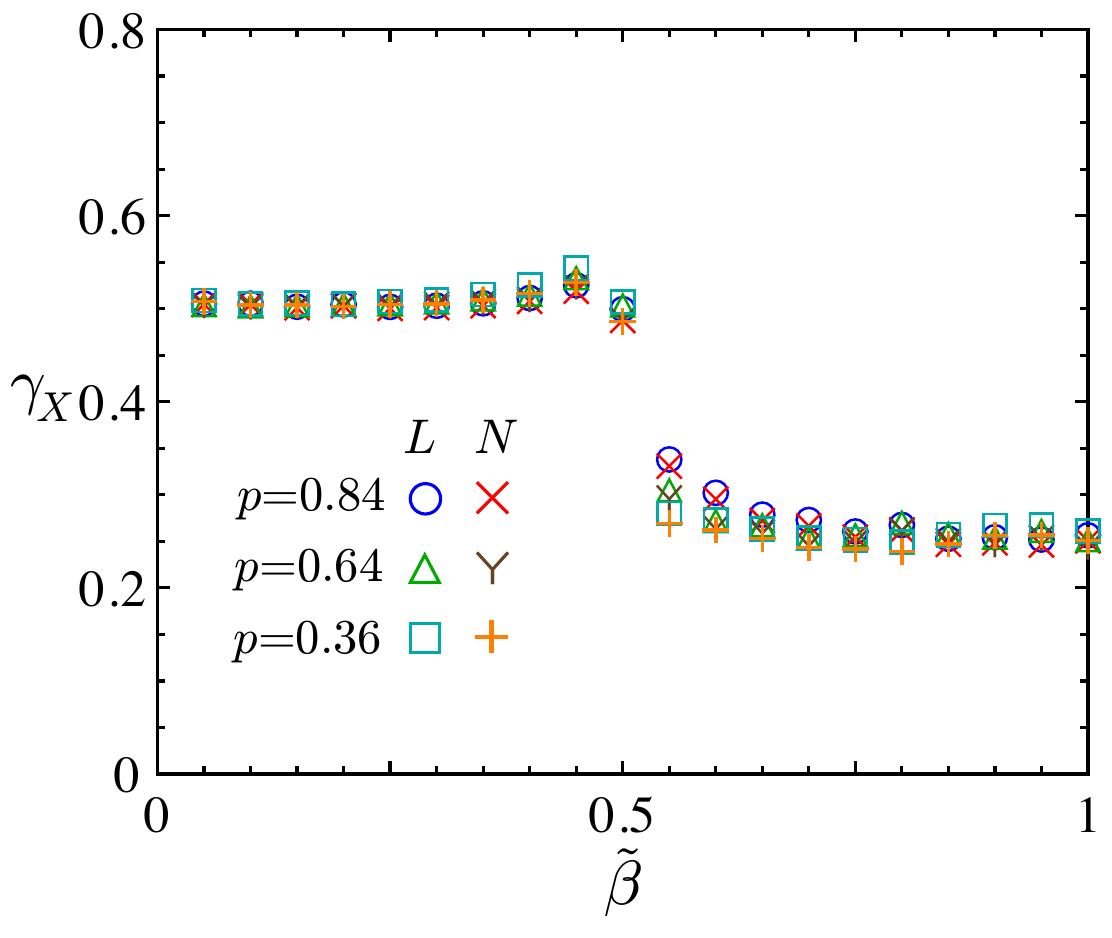}
\qquad
 \includegraphics[width=0.42\textwidth]{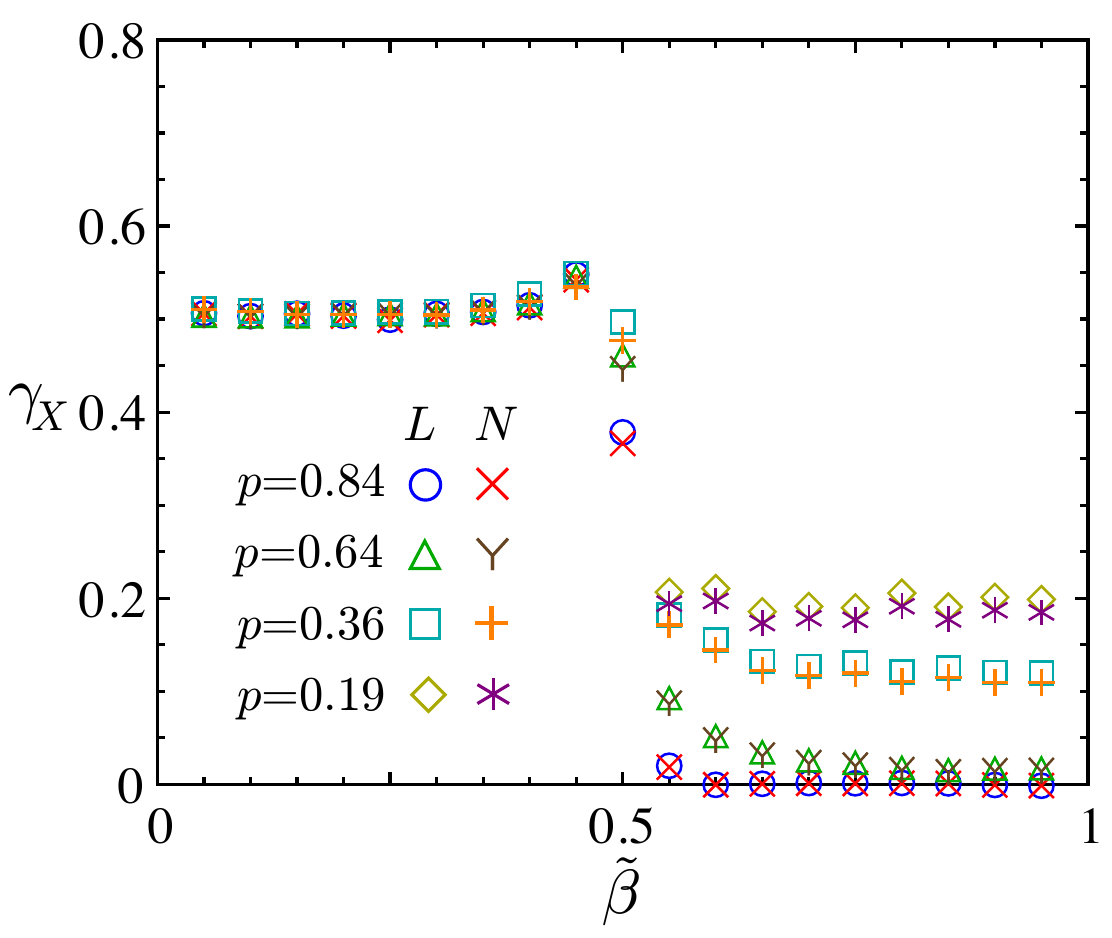}
\caption{Exponents $\gamma_X$ for parallel dynamics (left) and
backward dynamics (right).
$\beta$ has been rescaled so that $\tilde{\beta}=0,1/2,1$ corresponds to 
 $\beta =0,\beta_c,1$, respectively.
\label{fig-exponents}}
\end{center}
\vspace{-6mm}
\end{figure}

Fig.~\ref{fig-length} shows the time-dependence of the average
system length obtained by Monte Carlo simulations.  As one can observe in
these log-log plots, the slopes depend both on the update type and the
location on the critical line (curved part $\beta<\beta_c$ or straight
part $\beta>\beta_c$).  Fig.~\ref{fig-exponents} also provides
simulation results of the exponents.
Depending on the update rule, the exponents have different values:
\begin{eqnarray}\label{eq:gamma}
\text{parallel:}&&\qquad\gamma_X =
\begin{cases} 1/2 &   (\text{for }\beta<\beta_c) \\ 
1/4   &    (\text{for }\beta>\beta_c) 
\end{cases} \\
\text{backward:}&& \qquad
\gamma_X =
\begin{cases} 1/2 &  (\text{for }\beta<\beta_c), \\ 
g(p) &  (\text{for }\beta>\beta_c, p<p_c) \\
   0 &  (\text{for }\beta>\beta_c,p>p_c) 
\end{cases}   
\label{gammab}
\end{eqnarray}
with some function $g(p)\in (0,1/4)$, whose explicit form is not
known.  The nonuniversal behavior \eqref{gammab} and the existence of
the critical point $p_c$ for the backward case have been tested by
simulations ($t\lesssim 10^6$, averaged over up to $10^6$ samples
\cite{bib:AS4}).  We think that this is the most reasonable
interpretation, although one could consider other scenarios on
the straight part of the critical line $\alpha=\alpha_c,
\beta>\beta_c $ for the backward case,  e.g.
the average length that always
converges, but with extremely slow relaxation to the stationary length 
for small $p$.

%%%%%%%%%%%%%%%%%%%%%%%%%%%%%%%%%%%%%%%%%%%
%%%%%%%%%%%%%%%%%%%%%%%%%%%%%%%%%%%%%%%%%%%

%%%%%%%%%%%%%%%%%%%%%%%%%%%%%%%%%%%%%
%%%%%%%%%%%%%%%%%%%%%%%%%%%%%%%%%%%%%
 
%%%%%%%%%%%%%%%%%%%%%%%%%%%%%%%%%%%%%%%%%%%%%%%%%%%%%%%%%%%%%%%%%%%%%%%%%%%%%
\section{Model extensions}\label{sec:extensions}

%%%%%%%%%%%%%%%%%%%%%%%%%%%%%%%%%%%%%%%%%%%%%%
\subsection{EQP with Langmuir Kinetics}\label{sec:EQPLK}

The TASEP has been extended by including Langmuir Kinetics (TASEP-LK),
which is relevant for applications in biology and has a rich phase
diagram \cite{bib:PFF,bib:EJS}.  In a similar way we have combined
the parallel EQP \cite{bib:AY} with Langmuir kinetics
(EQP-LK).

\begin{figure}
\begin{center}
\includegraphics[width=0.75\textwidth]{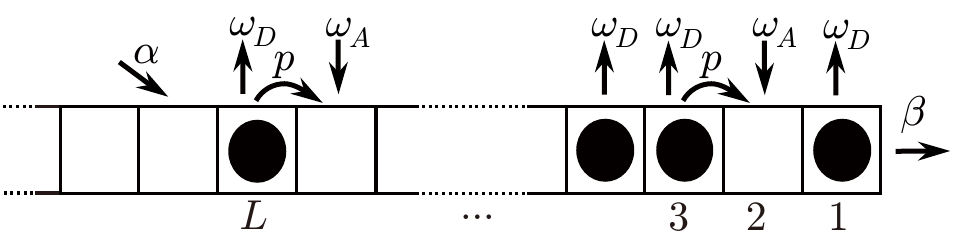}
\vspace{-3mm}
\end{center}
\caption{Exclusive queueing process with Langmuir kinetics.}
\label{fig:EQPLK} 
\end{figure}

In the presence of  Langmuir kinetics, particles in
the bulk are detached with probability $\omega_D$, and for each empty
site $j(\le L)$ a particle is attached with probability $\omega_A$
(see Fig.~\ref{fig:EQPLK}).
As in the TASEP-LK \cite{bib:PFF,bib:EJS}, 
the attachment and detachment probabilities are  
scaled as $\omega_A = \Omega_A/L$ and $\omega_D = \Omega_D/L$ which
leads to a competition between bulk and boundary dynamics. 
We do not impose attachment and detachment when the system length is $0$.
Note that, in contrast to the TASEP-LK, the system length $L_t$ of the
EQP-LK varies, and thus probabilities $\omega_A, \omega_D$ depend on
the current state.

In each time step, first the configuration is updated according to the
rule of the EQP with parallel update. Then the Langmuir kinetics is
applied. This defines the EQP-LK with parameters
$(p,\alpha,\beta,\Omega_A,\Omega_D)$, which reduces to the EQP for 
$\Omega_A=\Omega_D=0 $.  
The EQP-LK can be interpreted as an effective model for interacting
queues where the attachment and detachment corresponds to customers
changing from and to other queues, respectively. 
Thus the other queues are considered to act as reservoir for
the EQP-LK.

%%%%%%%%%%%%%%%%%%%%%%%%%%%%%%%%%%%%%
%%%%%%%%%%%%%%%%%%%%%%%%%%%%%%%%%%%%%

\begin{figure}
\begin{center}
\includegraphics[width=0.9\columnwidth]{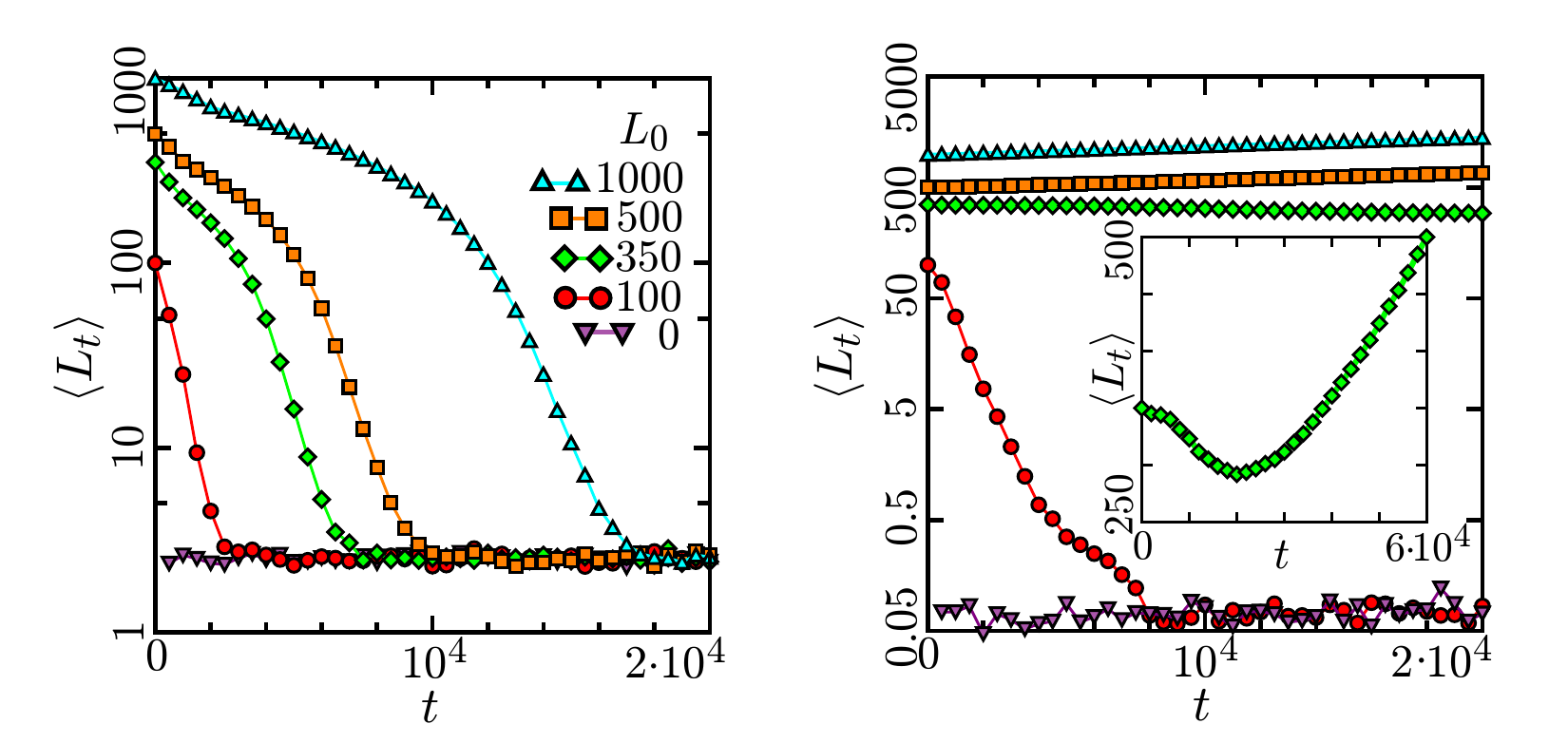}
\end{center}
\vspace{-3mm}
\caption{
  Behavior of $\langle L_t\rangle$ starting from various initial
  lengths, for
  $(p,\alpha,\beta,\Omega_A,\Omega_D)=(0.8,0.2,0.2,0.2,0.35)$ (left)
  and $(p,\alpha,\beta,\Omega_A,\Omega_D)=(0.8,0.3,0.2,0.1,0.9)$
  (right).  We have set the initial density as
  $\Omega_A/(\Omega_D+\Omega_A)$, and averaged over $10^3$ samples. In
  (a), all the average lengths with initial lengths
  $L_0=0,100,350,500,1000 $ converge to a stationary value.  In (b),
  we find that the behavior of the average length depends on the initial
  length.  Furthermore in the inset of (b), the length exhibits
  non-monotonic behavior. }
\label{fig:ergo-nonergo} 
\end{figure}

\begin{figure}
\begin{center}
\includegraphics[width=0.9\textwidth]{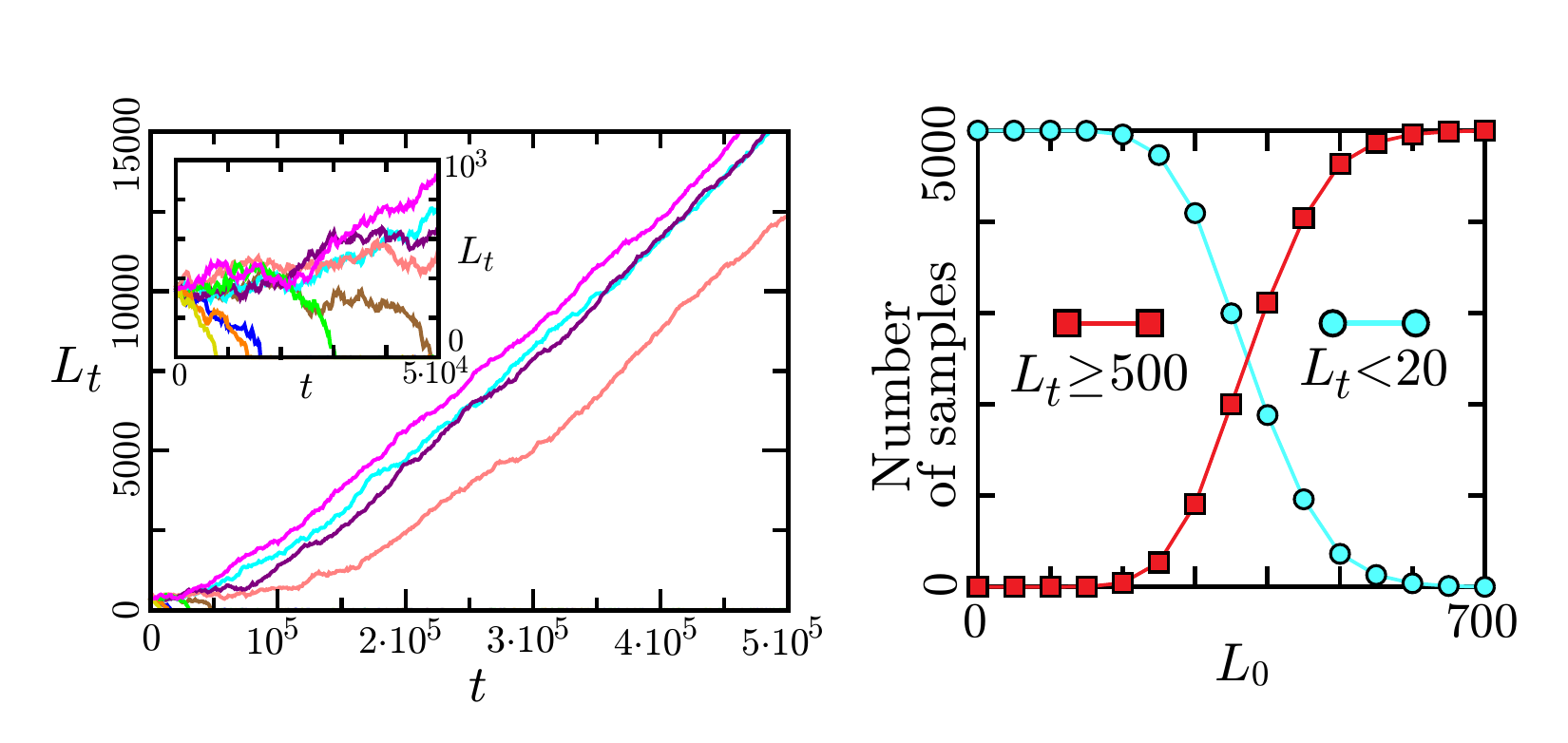}
\end{center}
\vspace{-3mm}
\caption{\label{fig:L0350}
  (Left) Behavior of $L_t$ for 9 individual samples.  The parameters
  were set as $(p,\alpha, \beta,\Omega_A, \Omega_D) =
  (0.8,0.3,0.2,0.1, 0.9)$, and every sample started from the length
  $L_0=350$ and the density $ \Omega_A/(\Omega_A+\Omega_D)=0.9 $.
  (Right) Distribution of the system length $L_t$ at $t=2\cdot 10^5 $
  obtained from $5\cdot 10^3$ simulation samples with the same
  parameter setting and various initial lengths $L_0$.  }
\end{figure}

Preliminary studies have found that the EQP-LK has surprising
properties \cite{Schultens,Borghardt,bib:inprep}.  It shows a strong
dependence of the behavior on the initial condition, see
Fig.~\ref{fig:ergo-nonergo}.  In fact, in a certain parameter region,
some samples appear to converge to a finite length whereas other
samples appear to diverge (within the simulation time), see
Fig.~\ref{fig:L0350}.  The percentage of apparently divergent samples
depends strongly on the initial length $L_0=L_{t=0}$ of the queue. It
is very small for small $L_0$ and becomes large for large $L_0$.

This surprising behavior is related to the length dependence of the
attachment and detachment probabilities. Once the queue has become
short it is difficult to escape from $L_t=0$ after reaching $L_t=0$
since the detachment rate is large.  Thus, starting from a short queue
$L_0<L^*$, the length tends to remain finite.  On the other hand,
starting from long queues $L_0>L^* $ the chain tends to grow further
because the detachment rate is small.  These two observations
contradict the general theory of Markov processes.  However, we can
interpret the convergence from short queues as ``quasi stationary,''
and it is required very long time to reach e.g. $ L=500$, which is
probably impossible to realize in our computer environment, see some
observations on the first passage time \cite{bib:inprep}.  In this
sense, the ergodicity of the EQP-LK is {\it effectively} broken.  In
other words, there is a very high maximum of the ``potential'' in a
certain point $L=L^*$.  This is opposite property to a microtubule
model \cite{bib:MRF}, where the length can be regulated around $L^*$.

%%%%%%%%%%%%%%%%%%%%%%%%%%%%%%%%%%%

%%%%%%%%%%%%%%%%%%%%%%%%%%%%

%%%%%%%%%%%%%%%%%%%%%%%%%%%%
%%%%%%%%%%%%%%%%%%%%%%%%%%%%

%%%%%%%%%%%%%%%%%%%%%%%%%%%%%%%%%%%%%%%%%%%%%%
\subsection{Multi-chain EQPs}

The EQP-LK is a single-chain queueing process that can be interpreted
as an effective model for a multi-chain process. The interaction
between the chains through exchange of particles is modeled through
Langmuir kinetics which allows the change of the particle number
within the bulk of the queue.

\begin{figure}
\begin{center}
 \includegraphics[width=0.7\columnwidth]{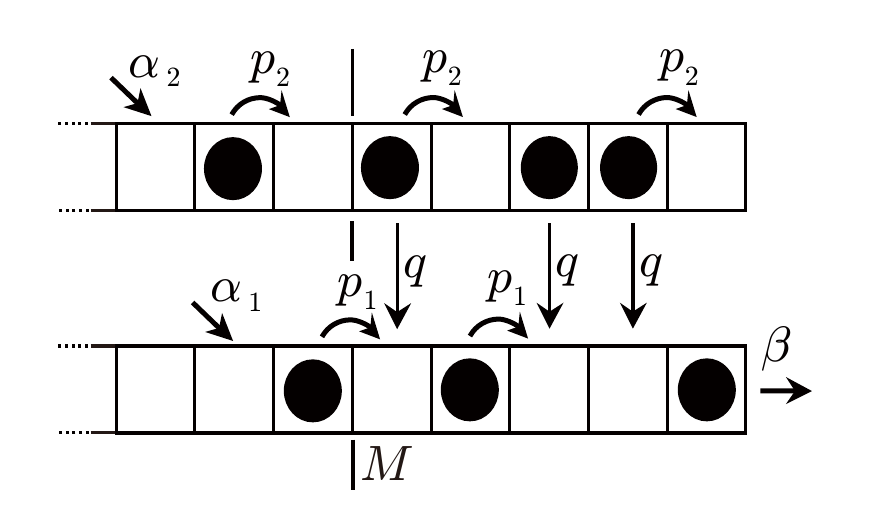}
\end{center}
\vspace{-5mm}
\caption{2-chain EQP for a highway bottleneck. The bulk dynamics
of each chain is that of the standard EQP. In a finite region
($j=1,2,\ldots,M$) 
particle changes from chain 2 (top) to chain 1 (bottom) are allowed.}
\label{fig:bottleneck} 
\end{figure}

 Of course, the EQP can also be extended to a genuine multi-chain
model. For applications, 2-chain models are of special interest.
As a model for a bottleneck on a highway, a configuration as
in Fig.~\ref{fig:bottleneck} can be used. Here particles can
only change with probability $q$ from chain 2 to chain 1 on 
sites $j=1,\ldots,M$ where $M$ is a fixed parameter. For sites
$j>M$ such a change is not allowed. Otherwise, the bulk and
input dynamics (with parameters $p_1,p_2$ and $\alpha_1,\alpha_2$,
respectively) are identical to that of a single EQP. However,
only particles at position $j=1$ of chain 1 are serviced.

%%%%%%%%%%%%%%%%%%%%%%%%%%%%
%%%%%%%%%%%%%%%%%%%%%%%%%%%%

%%%%%%%%%%%%%%%%%%%%%%%%%%
\section{Applications and related models}
\label{sec-applications}

Originally, queueing theory was developed mainly for applications
in telecommunications. Nowadays, however, it is a standard approach
in various fields, ranging from supply chains \cite{Reffactory} to
traffic flow and biology.

The most natural application of the EQP is queueing of pedestrians,
e.g. at a supermarket checkout. This can be generalized in a
straightforward way to multiple queues where customers might jump from
one queue to another.  A generalization where the probability of
moving depends on the gap to the next customer in front was studied in
\cite{behlau}. This might be realistic e.g.  for queues at an airport
check-in where the passingers have to pick up their luggage when
moving forward. Since this is uncomfortable they typically wait until
a critical gap to the preceding passenger has opened.

One advantage of the EQP and its relatives for applications is that
it is an intrinsically microscopic model where the different ``units''
can be distinguished. Therefore they can have different properties
(e.g. average velocities) in a natural way.

In the context of vehicular traffic, various queueing based models
have been proposed, e.g.
\cite{RefHeid,EissfeldGW03,Helbing03,Vandaele,RefCaca,RefMacG}.  Even
the cell transmission model \cite{Daganzo} might be interpreted as
queueing model. Often these models are used to study traffic on
networks where links correspond to roads or road sections and nodes to
intersections which are characterized by a service rate.  In contrast,
in the model developed in \cite{RefCaca} the trajectories of the
vehicles are related to a M/M/1 queue by identifying space in the
traffic model with time in the queueing model.

Other variants of the TASEP on a lattice of varying length have been
proposed as applications to biological systems. In
\cite{Evans1,Evans2,Evans3,bib:DMP} the dynamically extending
exclusion process (DEEP) has been introduced as a model for fungal
growth.  In the DEEP not all particles are removed from the system as
they reach the end, but with some probability form a new lattice site.
In contrast to the EQP, the DEEP has no mechanism for reducing the
system length and therefore the length of the system is always
diverging. Microtubules \cite{ref:howard} are the analogues of highways
in cells. However, their lengths are not constant, but changes
dynamically. The mechanism of length regulation of microtubules has
been investigated in \cite{bib:JEK,bib:MRF} using a variant of the
TASEP where the first (output) site can be removed or attached under
certain conditions. Similar models have been used to describe
bacterial flagellar growth \cite{bib:SchS}.  In \cite{bib:MRF}, a
condition on parameters for the convergence of a simplified model of
mictrotubules is discussed.  This can be also rigorously derived by
constructing a stationary measure in a similar form to that of the EQP
\eqref{arrmpf2} \cite{bib:ALS}.

%%%%%%%%%%%%%%%%%%%%%%%%%%%%%%%%%%%%%%%%%%%%%%%%%%%%%%%%%%%%%%%%%%%%%%%%%%%%%
\section{Discussion}
\label{sec:discussion}

The Exclusive Queueing Process (EQP) can be considered as a minimal
model of pedestrian queues which takes into account the internal
dynamics of the queue.  We have found that the EQP has a rich phase
diagram.  Surprisingly, it shows a strong dependence of its critical
properties on the update scheme.  This is rather different from the
TASEP with a fixed system length.  The order of the phase transition
between the diverging and converging phases can also be different.

Besides application to pedestrian queues and vehicular traffic, variants 
of the EQP have interesting applications to biological processes
like fungal growth and microtubule length regulation.  
We expect that transport models with varying system lengths will show
many other interesting phenomena.

%%%%%%%%%%%%%%%%%%%%%%%%%%%%%%%%%%%%%%%%%%%%%%%%%%%%%%%%%%%%%%%%%%%%%%%%%%%%%

%
\section*{Acknowledgment}
AS is partially supported by Deutsche Forschungsgemeinschaft (DFG)
under grant ``Scha 636/8-1''.  The authors are grateful to Christoph
Behlau, Christian Borghardt, Alex T. L\"uck, Ludger Santen, Christoph
Schultens and Daichi Yanagisawa for collaboration works which we
partially reviewed in this article. The authors also thank Martin R.
Evans, Kirone Mallick and Gunter Sch\"utz for useful discussions.

%%%%%%%%%%%%%%%%%%%%%%%%%%%%%%%%%%%%%%%%%%%%%%%%%%%%%%%%%%%%%%%%%%%%%%%%%%%%%

%\section*{References}

\end{document}